\documentclass[12pt,a4]{article} %[aps,prx,preprint,groupedaddress]{revtex4-2}
\usepackage{graphicx}
\usepackage{subfigure}
\usepackage{color}
\usepackage{silence}
\graphicspath{{figs/}}
\usepackage{hyperref}

\title{Aster Swarming by Collective Mechanics of Dyneins and Kinesins}%}A Model of Spontaneous Microtubule Aster Vortexing in Cellular Confinement with Cortical Dyneins and Diffusible Kinesins} 
\author{Neha Khetan and Chaitanya A. Athale,\\ Div. of Biology, IISER Pune, Dr. Homi Bhabha Road,\\ Pashan, Pune 411008, India.\\ Email: \href{mailto:cathale@iiserpune.ac.in}{cathale@iiserpune.ac.in}.}

\begin{document}
\maketitle

\section{Abstract}
Microtubule (MT) radial arrays or asters establish the internal topology of a cell by interacting with organelles and molecular motors. We proceed to understand the general pattern forming potential of aster-motor systems using a computational model of multiple MT asters interacting with motors in a cellular confinement. In this model dynein motors are attached to the cell cortex and plus-ended motors resembling kinesin-5 diffuse in the cell interior. The introduction of `noise' in the form of MT length fluctuations spontaneously results in the emergence of coordinated, achiral vortex-like rotation of asters. The coherence and persistence of rotation requires a threshold density of both cortical dyneins and coupling kinesins, while the onset of rotation if diffusion-limited with relation to cortical dynein mobility. The coordinated rotational motion arises due to the resolution of the `tug-of-war' of the rotational component due to cortical motors by `noise' in the form of MT dynamic instability, and such transient symmetry breaking is amplified by local coupling by kinesin-5 complexes. The lack of widespread aster rotation across cell types suggests biophysical mechanisms that suppress such intrinsic dynamics may have evolved. This model is analogous to more general models of locally coupled self-propelled particles (SPP) that spontaneously undergo collective transport in presence of `noise' that have been invoked to explain swarming in birds and fish. However, the aster-motor system is distinct from SPP models with regard to particle density and `noise' dependence, providing a set of experimentally testable predictions for a novel sub-cellular pattern forming system.

\section{Popular summary}{\label{summary}} %word limit 250
Radial microtubule (MT) arrays or asters determine the geometry of animal cells and determine their plane of division. The positioning and transport of asters in cells has been addressed in previous work based on multiple force-generators. We develop a computational model of multiple asters in a cell that combines previously reported forces and examines the potential for collective motion. Forces are generated in our model by either the walking activity of biomolecular motors or MT bending mechanics. Motors of two kinds are included - minus-end directed dyneins localized on the cell boundary and coupling kinesins in the cytoplasm. The model geometry is based on experimental reports of animal cells ranging from epithelial cells to one-celled embryos. Simulations result in spontaneous random motion of the asters when `noise' due to MT length dynamics is added. The rotational motion is equally likely to result in clockwise and anti-clockwise rotations, due to the random nature of the initiation of motility. Additionally, we find the collective motion requires the dyneins to be mobile on the cell boundary, with diffusive mixing essential for pattern formation. This model predicts a form of collective motion, similar to that observed during mitosis in embryonic development. Our model predicts such effects are intrinsic to cells, and mechanisms to suppress such swarming motility of asters might have evolved that are yet to be tested in experiment. Aster swarming based on local interaction of components and `noise' that cannot be predicted from the components alone, is analogous to collective schooling of fish and swarming of birds, which lack an explicit leader. This suggests it belongs to a general class of self-propelled particle (SPP) models, with differences in specific features from such general models. 
%Physical Review X requires authors to submit a nontechnical summary that conveys the context, the essential message(s), and the significance of the work to all readers. The summary should be concise (approximately 250 words), readable, objective, and have broad appeal. Please avoid including mathematical expressions.
%% ===============================================================

\section{Introduction}{\label{intro}}
Self-organized pattern formation %resulting in the emergence of order from simpler interactions of many individual components at an underlying level of complexity 
is observed almost universally in biological systems %\cite{Camazine2003,smith1998shaping}. Such patterns 
ranging in scales from large scale structures of swarming birds and fish \cite{camazine1990}, through cells undergoing collective migration patterns\cite{friedl1995,szabo2006,reffay2011,tambe2011,rausch2013,soumya2015}, to single-cell polarization by reaction-diffusion networks of proteins \cite{Altschuler2008,asano2008,taniguchi2013}. %Cytoskeletal filaments and molecular motors are critical regulators of mechanics of cellular shape, size and movement.  
The self-organized patterns of the mechanical elements of the cell, the cytoskeleton and molecular motors are particularly distinct, arising as they do from purely mechanical interactions. Of these the most distinct are the {\it in vitro} patterns reported from the reconstitution of ATP containing mixtures of microtubules (MT) and motors \cite{nedelec1997,nedelec2001,surrey2001,Sumino2012} and actin and myosin \cite{Backouche2006,Ennomani2016}. MT-motor activity in a circular boundary has been shown to break symmetry and result in vortex like motility \cite{nedelec1997,Suzuki2017}. Evidence that such vortices are not just restricted to minimal {\it in vitro} reconstituted systems, has come from {\it in vivo} studies demonstrating MT-motor driven cytoplasmic streaming in {\it Caenorhabditis elegans} embryos \cite{Shinar2011} and {\it Drosophila} ooocytes \cite{lu2016} during development and the cells of the plant {\it Chara} \cite{vandemeent2008}. However, the range of motility patterns seen in these structures are specific to the geometry of the cell type and specific mix of motors. In order to understand the general principles of such pattern formation, theoretical models that take into consideration a wider range of cell and filament geometries are required.

A general model of self propelled particle (SPP) motion describing the emergence of collective motion or swarming based on local coupling interactions and `noise' has been described by Vicsek et al. \cite{vicsek1995}. While this class of models minimally requires active particles with local interactions and `noise', they do not capture the polarity of filaments and motors, since active particles are typically considered to be point particles. MT filaments have a kinetic polarity of plus- and minus-ends, that determines the direction of motor activity- kinesins walk towards the plus-ends and dyneins towards the minus-ends \cite{howard2001}.  % {\it In vitro} reconstituted microtubule-motor systems are relatively ease of control and mechanistically better understood compared to whole cells, as a result of which they have been used in multiple studies demonstrating the emergence of higher-order patterns of  \cite{nedelec1997,Suzuki2017}. 
Detailed models of self-organized patterns of linear filaments have shown good agreement with experiments as seen with MTs in presence of kinesin \cite{nedelec1997,nedelec2001,surrey2001} or dynein \cite{Sumino2012} as well as actin with myosin activity \cite{Backouche2006,schaller2010,Ennomani2016}. However, inside most animal cells MTs have a characteristic orientation of minus-ends near the nucleus and plus-ends at the cell periphery, forming a mechanical positioning system for organelle transport, cell polarization and cell division. This characteristic organization of MTs is determined by microtubule organizing centers (MTOCs) that serve as nucleation points forming radial arrays or asters. At the same time most studies of collective MT transport have used linear MTs. The effect of MT geometry on mobility is seen when comparing the outcome of these two kinds of MTs encountering a sheet of immobilized motors of either plus- or minus-ended type. While linear MTs undergo collective transport and glide, radial asters undergo a tug-of-war due to geometry as seen with kinesins for 1D doublets \cite{leduc2010} or dyneins transporting asters \cite{athale2014}. The effect of diffusible motors on aster movement is variable and determined by both the motor type - tetrameric kinesins and MT orientation. Pairs of asters coupled by kinesins bound to anti-parallel MTs will form bipolar spindle-like structures in simulations \cite{nedelec2002}, confirming the importance of kinesin-5 in spindle assembly seen in experiments \cite{walczak1998}. Such anti-parallel MTs from asters of neighboring spindles in syncytial embryos of {\it Drosophila} result in even spacing of spindle asters \cite{telley2012}. Parallel MTs on the other hand result in `zippering' by movement of motors on parallel MTs as seen in with MTOC asters during in mouse oocyte spindle assembly \cite{schuh2007,khetan2016}. Dyneins coalesce asters independent of whether MTs are parallel or anti-parallel and the effect is observed supernumary centrosome clustering due to dynein \cite{quintyne2005}. In many cells however, dyneins are immobilized in the cell cortex. Asters contacting these membrane anchored motors result in radial pulling forces acting on asters driving them to the cell center in cells with sizes comparable to the asters \cite{laan2012}. In multi-nucleate cells such as the filamentous fungus {\it Ashbya gossypii}, these astral-MT interactions with cortical dynein are essential for maintaining a regular spacing between nuclei \cite{Gibeaux2017}. Cortical dynein localization forms the mechanical basis of the asymmetric cell division of one-celled embryos of {\it C. elegans} \cite{grill2003} and the cortical density of  dynein has been shown to determine spindle oscillations \cite{Pecreaux2006}. This suggests mechanical interactions of MT asters with diffusible kinesin tetramers and cortical dyneins constitute a conserved mechanical modules across a wide variety of cell types. While a general theory for the mechanics of a single aster in a confined cellular geometry has been developed previously \cite{ma2014}, a model integrating motor localization seen in diverse cells with multiple asters is lacking. % suggesting their collective behaviour might deviate from that seen in previous work with linear MT filaments. The ability of such aster-like geometries to self-organize due to motor activity, remains to be explored. %A systematic examination of 

%The control over copy number of centrosomes that nucleate asters is an important regulatory process in animal cells, and dysregulation results in chromosome segregation defects as seen in some cancers \cite{hornick2008}. Additionally multiple aster positioning and dynamics are influenced by motors and MT dynamics as seen in mouse meiotic oocytes \cite{schuh2007,clift2015} and syncytial embryo spindles of {\it Drosophila} \cite{telley2012}. Thus, an improved understanding of the effect of collective mechanics of asters on spatial organization is  relevant to {\it in vivo} scenarios in development and disease.
%---------------------------------------------------------------------------------------------------
\begin{figure}[ht!]
	\begin{center}
		\includegraphics[width=0.95\textwidth]{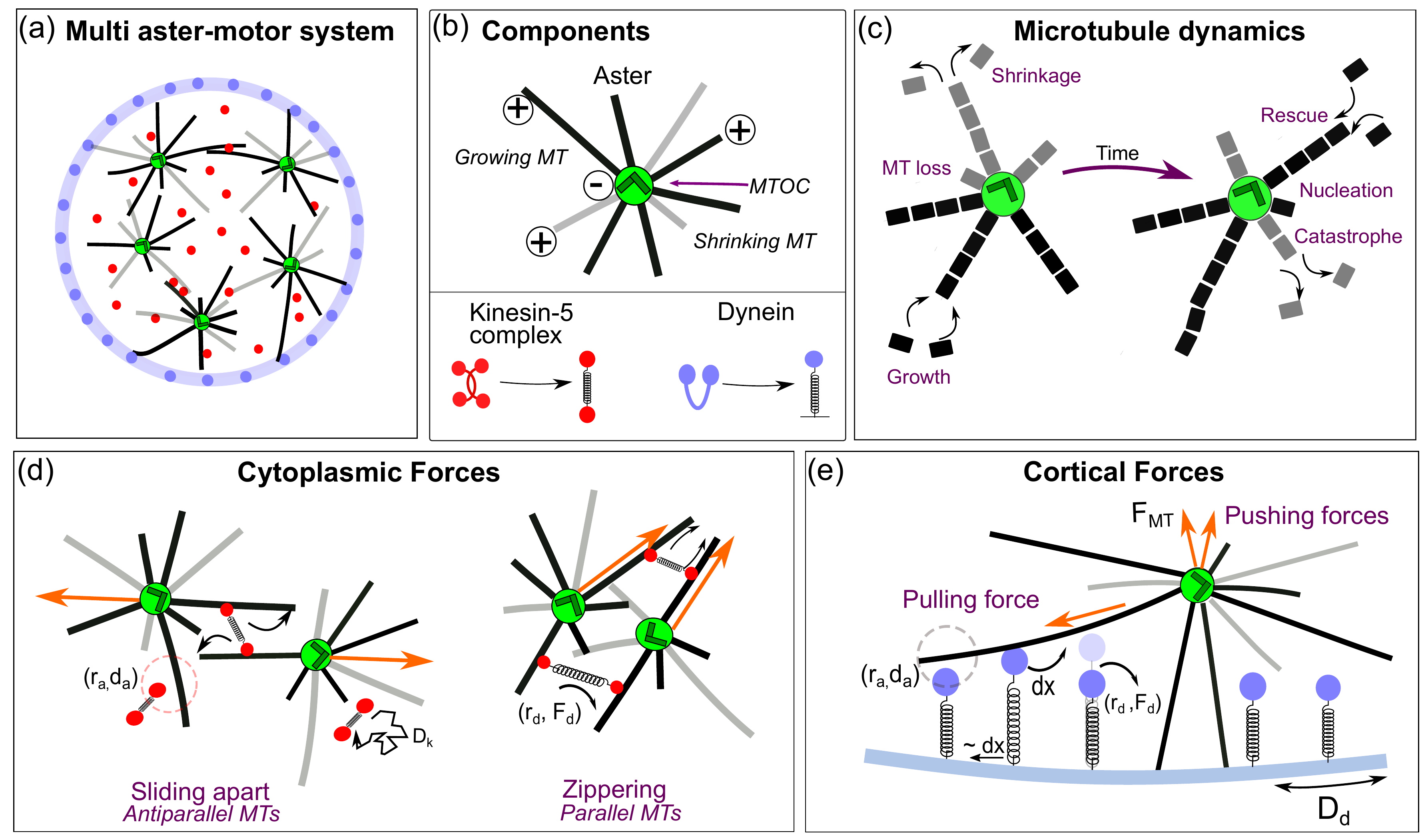}
		\caption{{\bf Model of aster-motor mechanics.}(a) The model consists of multiple asters and motors in confinement. (b) Aster MTs are radially nucleated with minus-ends of MTs embedded in the Microtubule (MT) organizing center (MTOC) and plus-ends out. MTs can switch between growing (black) and shrinking (gray) states. Motors are modeled as springs MT binding sites as seen in case of anchored dynein (blue) and  kinesin-5 (red) with two motor domains. (c) MTs dynamic instability is modeled with transitions between growth and shrinkage states, while MTs per aster remain constant. (d) In the cell interior unbound kinesins diffuse and and can generate forces when bound a pair of MTs, serving to separate or aggregate asters, based on MT orientation. (e) At the cortex asters are pushed inwards by MT bending forces ($F_{MT}$), while anchored dyneins when bound drive aster transport. Free dyneins diffuse along the cortex. %The polymerizing MT bends at the rigid cell cortex and generates inward pushing forces ($F_{MT}$).
Orange arrows: force on the asters. Black arrows: direction of motor movement.}
		\label{fig:asterschem} 
	\end{center}
\end{figure}
%---------------------------------------------------------------------------------------------------
Here, we have modeled the mechanics of a mixture of multiple MT asters acted on by dyneins at the cell cortex and kinesins in the cytoplasm, to examine the potential of this system for collective transport. %onsisting of semi-flexible filaments in a radial array around centrosomes. We 
We test the effect of the forces generated on asters arising from local coupling by cytoplasmic kinesins, gliding of MTs on the circular boundary lined with cortical dyneins and inward pushing due to MT mechanics in confinement and the role of stochastic MT polymerization dynamics. Our model demonstrates that while coupled mechanics alone results in local and uncorrelated aster motility, the addition of `noise' transforms it into coherent rotational motion. This emergence of coherent rotation or swarming of asters depends on cortical dyneins, kinesins and the MT stochasticity. %Additionally, similar to the general Vicsek type models, the system requires
The spontaneous coherent streaming motion of asters predicted by the model is discussed in the context of experimental evidence for the nature of dynein anchoring to the cell cortex. %The specific details of our model deviate from the simpler generalization, and aspects of diffusion limitation and density effects suggest a novel class of collective motion.

\section{Model}{\label{model}}
{\it Components and interactions:} We have modelled a multi-aster system in a cellular compartment in the over damped regime with mechanics and stochastic binding kinetics of kinesin and dynein (Figure \ref{fig:asterschem}), based on a previously developed computational agent-based model of MT-motor interactions \cite{nedelec2007}. MTs form a radial structures, asters, that interact with one another when they are cross-linked by molecular motors that bind to and walk on MTs. This mechanical coupling of asters at micrometer scales originates from motor-driven forces spanning a few nanometers. The net force experienced by a complex of coupled asters determines whether they are transported or static. Stochasticity in the model originates from three sources: MT length fluctuations, diffusion and the binding kinetics of motors. The mechanical interactions vary according to position either at (a) the cell cortex or (b) in the cytoplasm and are described separately.\\
(a) {\it Cell cortex:} Dynein-like motors are modeled as being bound to the cell cortex by their stalk domains and can bind to and walk on MT filaments resulting in aster transport along the cell boundary. Such movement is comparable to a `gliding assay' described in previous work \cite{athale2014,khetan2016,jain2019}, with a rotational component due to the circular geometry. MTs when bound to dyneins or non-specifically encountering the cell-membrane at the cortex, bend and generate an inward restoring force on the aster (Figure \ref{fig:asterschem}). In addition, MT-bound motors can also be dragged through the membrane. When multiple motors bind to oppositely oriented MTs of the aster a tug-of-war arises in the MT-motor system. Such tug-of-wars emerge from the radial geometry of asters, as previously described \cite{athale2014}. Dyneins that are not bound to MTs, can diffuse in the membrane.\\ 
(b) {\it Cytoplasm:}  In the cell interior, diffusible tetrameric kinesin motors are modeled based on the mitotic kinesin-5 complexes. They can bind MTs and walk towards the plus end and when bound to astral MTs from neighboring asters simultaneously, they produce forces on asters. The resultant force depends on MT orientation- parallel or anti-parallel, leading to either `coalescence' or `separation' of asters respectively (Figure \ref{fig:asterschem}). Model parameters are taken from experimental reports where possible (Table \ref{tab:simparams}).

{\it MT asters:} A microtubule organizing center (MTOC) is modeled to nucleate a finite number of MTs in a radial manner forming an aster. MTs are modelled as semi-flexible rods with a bending modulus $\kappa$ of 20 $N/m^{2}$, based on previous reports \cite{gittes1993}. When MTs at the cell boundary bend, they experience a restoring force based on beam-theory, $F_{bend}=\pi^2 \cdot \kappa/L^2$, where L is the filament length and $\kappa$ is the flexural rigidity of microtubules. Bending can occur either because the aster is held at the rigid cell boundary, or MTs are bound to multiple dyneins. MT lengths ($L_{MT}$) are modelled as either uniform and of fixed lengths (stabilized) or fluctuating (dynamic instability) with a mean length. We use the mean length of asters from measurements made on {\it Xenopus} oocytes \cite{verde1992} of $\langle L_{MT} \rangle=$ 4.25 $\mu m$. The length dynamics are described by the frequencies of catastrophe ($f_c$) and rescue ($f_r$) and velocities of growth ($v_g$) and shrinkage ($v_s$) with values taken from those reported for {\it Xenopus laevis} oocyte extracts \cite{verde1992,athale2008}: $f_c = 0.049$ 1/s and $f_r= 0.0048$ 1/s, $v_g = 0.196$ $\mu m/s$ and $v_s = 0.325$ $\mu m/s$. The values of f$_{c}$ and f$_{r}$ are 0.049 1/s and 0.0048 1/s respectively in all calculations unless mentioned.
%------------------------------------------------------------------------------------------------------------------------------------------------------
\begin{figure}[ht!]
	\begin{center}
		\includegraphics[width=0.8\textwidth]{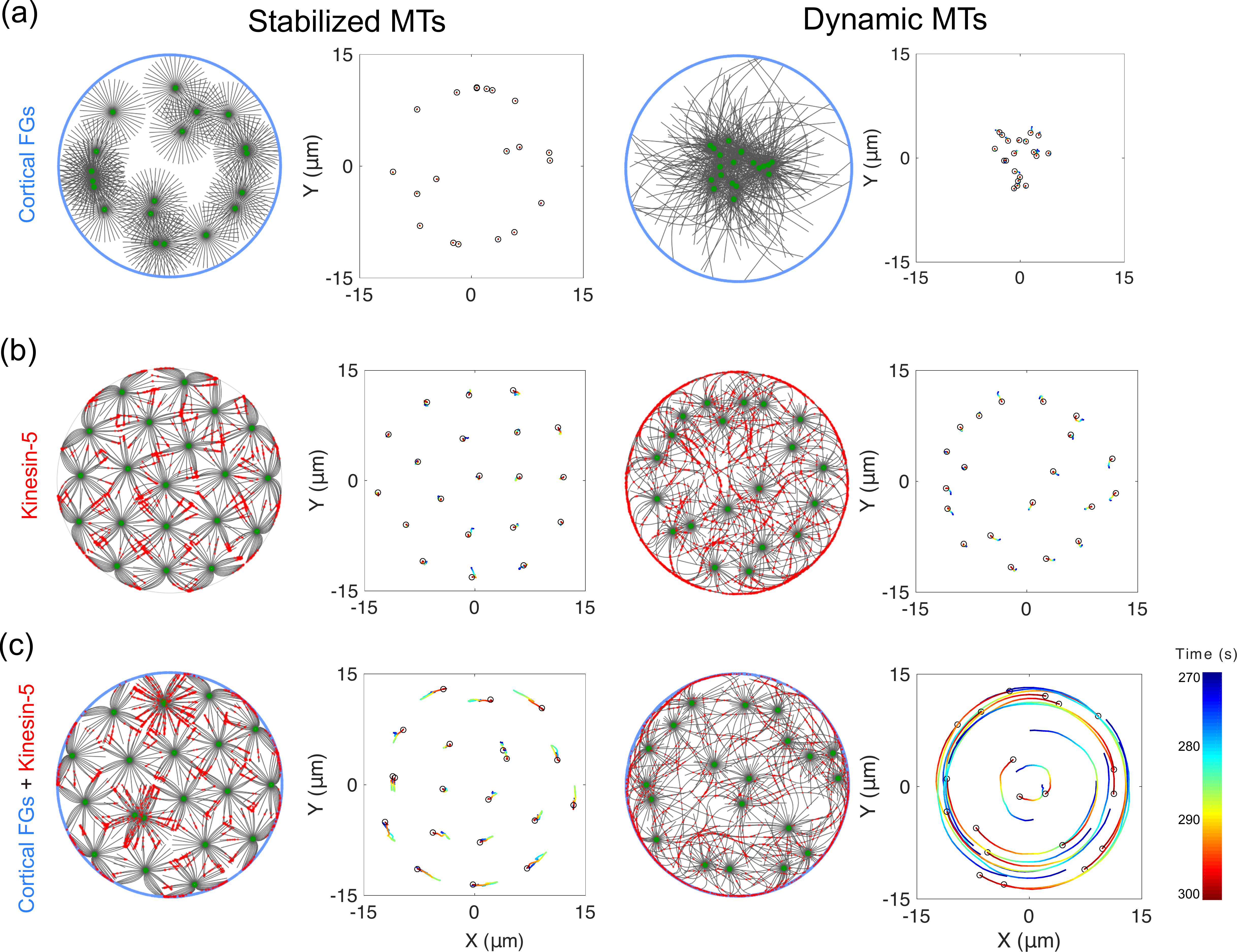}
		\caption{{\bf Spontaneous aster rotation.} Asters with MTs that were either ({\it left}) of fixed length with MT flux J = 0 or  ({\it right}) stochastically fluctuating in length with J = -0.3 $\mu$m/s. Simulations were run in presence of one of the motor combinations: (a) diffusible cortical dyneins (blue circles) (b) cytoplasmic tetrameric kinesin-5 motors (red circles) or (c) dyneins and kinesins. The last 30 s of the trajectories of aster centers of the corresponding simulations are plotted (the colorbar represents time). Kinesin density $10$ motors/$\mu m^2$, cortical dynein density $10^2$ motors/$\mu m$, $\eta= 0.05$  $N\cdot s/m^2$ , $N_{a}=$ 20, $N_{MT}$ per aster = 40, $R_{cell}=15$ $\mu$m and total time 300 s. Video S1 represents the time-series.} %for stabilized MTs and  for dynamic MTs.}
		%(Video \ref{vid:patterns})
		\label{fig:minimalRotationSystem}
	\end{center}	
\end{figure}
%------------------------------------------------------------------------------------------------------------------------------------------------------
%Cortical and cytoplasmic coupling 

{\it Motor mechanics:} The motors modeled are of two kinds, cortical dyneins and diffusible kinesin-5 complexes. Cortical dyneins are modelled as single walking motors with a spring-like stalk domain (Figure \ref{fig:asterschem}), the spring constant $k_{d}$ of which determines the stretch force $F_s=-k_d$ $\Delta h$ for a $\Delta h$ change in stalk length \cite{athale2014, khetan2016}. Bound motors are modeled as discrete steppers with step size determined by the load. A load-free motor is modeled to take constant sized steps while increasing opposing forces reduces the step size in a piecewise manner based a previously described model \cite{athale2014,khetan2016}. Additionally, the anchored dyneins attached to the MT are dragged along the cell-boundary based on the stretch on motor due to the opposing force originating from the spring stretch force ($F_s$). The detachment and stall forces are identical ($f_{d} =f_{s}$) for dynein. Dyneins not attached to the MT diffuse in the membrane with an effective diffusion coefficient $D_d$, the value of which is varied (Table \ref{tab:simparams}).
Kinesin-5 motor complexes are modeled as two motor heads joined by a Hookean spring like connector representing the stalk (Figure \ref{fig:asterschem}), based on the stalk-linked dimer-of-dimers structure of this motor, that walks on two MTs simultaneously, thus coupling them \cite{walczak1998}. The complexes are diffusible throughout the interior of the cell with an effective diffusion coefficient $D_k$, the value of which we take to be 20 $\mu$m$^2$/s based on typical cytosolic proteins (Table \ref{tab:simparams}). Motors stochastically bind to MTs within a distance $d_{a}$ based on an attachment rate of $r_{a}$. The second motor-domain can similarly bind another MT, independent of the MT orientation. A bound motor walks on the MT at constant velocity $v=v_0$, if there is no opposing load acting on it. In the presence of an opposing load, the velocity becomes  $v=v_{0} \cdot ( 1 - \frac{f_{\parallel}}{f_{s}})$, where $f_{s}$ is the stall force and $f_{\parallel}$ is the projection of the extension force ($f_{ex}$) along the MT. In the case of the multi-aster system, a kinesin complex bound to two filaments simultaneously, will experience a load, resulting in a restoring force that drives aster movement. This is based on similar collective motor mechanics models used to model kinesin-5 in previous work \cite{loughlin2010}. Both dynein and kinesin-5 motors detach based on Kramers theory \cite{kramers1940} with a rate $r_d = r_{0} \cdot e^{| f_{ex}| /f_{d} }$ where $r_{0}$ is the load-free basal detachment rate and $f_{d}$ is the detachment force. When a kinesin-5 motor walks to the end of the MT it is modelled to detach immediately \cite{korneev2007, loughlin2010}. 

{\it MT dynamics:} MT dynamics is modeled based on the two state model of growth and shrinkage, associated with 4 parameters, the filament growth velocity $v_g$ and shrinkage velocity is $v_s$ and two transition frequencies between the two states: $f_{cat}$ the frequency of catastrophe (growth to shrinkage transition) and $f_{res}$ the frequency of rescue (shrinkage to growth transition) \cite{hill1987,Bayley:1989aa}. Work by Verde et al. \cite{verde1992} demonstrated how these parameters relate to mean MT length (L$_{MT}$) as:
\begin{equation}
L_{MT}  = \frac{v_{g} \cdot v_{s}}{v_{s} \cdot f_{c} - v_{g} \cdot f_{r}}.
\label{eq:mtdynlen} 
\end{equation}
and. The related variable of flux in MT lengths $J$ is then calculated by:
\begin{equation}
J  = \frac{v_{g} \cdot f_{r} - v_{s} \cdot f_{c}}{f_{c} + f_{r}}
\label{eq:mtdynflux} 
\end{equation} 
which determines whether the length of MTs on an average is in the `bounded state' ($J<0$) or unbounded state ($J>0$) or not dynamic ($J=0$). 
%where, L$_{MT}$ is the mean MT length at steady state, v$_{g}$ and v$_{s}$ are the growth and shrinking rates respectively, f$_{c}$ and f$_{r}$ are the frequency of catastrophe and rescue respectively. s in Equation \ref{eq:mtdynflux}. The frequency of catastrophe (f$_{c}$), frequency of rescue (f$_{r}$) was estimated by solving the two equations (Equation \ref{eq:mtdynlen} and \ref{eq:mtdynflux}) simultaneously. The velocity of growth (v$_{g}$), velocity of shrinkage (v$_{s}$) and mean MT length (L$_{MT}$) was held constant on decreasing the MT flux values by order of magnitudes. 

Thus the position of asters is determined by the net force that results from all these sources (Figure \ref{fig:asterschem}) i.e. inward pushing force due to MT bending at the cortex ($F_{MT}$), a force that pulls the asters to the cell boundary due to dyneins and separating or `zippering' forces in the cytoplasm due to kinesin-5 motility when bound to pairs of parallel or anti-parallel MTs respectively, with MT dynamics as a major source of stochasticity (Figure \ref{fig:asterschem}).

\begin{table}
	\caption{{\bf Model parameters.} The parameters that determine the mechanics and dynamics of motors are taken from experimental measurements reported in literature, and where missing estimated.}
	\label{tab:simparams}
	\begin{tabular}{ {l}{l}{l}{l}{l}} %{p{1.8cm} p{3.5cm} p{1.8cm} p{1.8cm} p{1cm}}
		\hline
		{\bf Symbol} & {\bf Parameter} & {\bf Dynein} & {\bf Kinesin-5}  & {\bf Reference}  \\    
		\hline 
		\\     
		$D$       & Diffusion coefficient  &  0-100 $\mu$m$^2$/s   & 20  $\mu$m$^2$/s   & \cite{wu2018membranewaves} \\
		$v_{0}$   & Motor velocity                 &  2                      & 0.04     $\mu$m/s              &  \cite{athale2014,loughlin2010} \\ 
		$d_{a}$   &  Attachment distance    &  0.02  $\mu$m	     	   & 0.05       $\mu$m    		  & \cite{athale2014,khetan2016,loughlin2010}\\             
		$r_{a}$   &    Attachment rate      &  12     $s^{-1}$          	   & 2.5          $s^{-1}$       		  &     ''   \\          
		$k$       &   Linker strength       &  100          pN/$\mu$m  	   & 100    pN/$\mu$m       		  & ''  \\	
		$f_s$     & Stall force                    &  1.75                   & 5        pN                    &    ''     \\                                              
		$r^{'}_{d}$     &  Basal detachment rate  &  1	  $s^{-1}$  	   & 0.05         $s^{-1}$  		  &  '' \\                   
		$r^{'}_{d,end}$ & Basal end-detachment rate   &  1 	    $s^{-1}$ 	   & immediate        		  & '' \\ 
		$f_d$	          &  Detachment force     &  0.5 pN            	   & 1.6          pN	  & \cite{korneev2007}, this study \\
		\\
		\hline
	\end{tabular}
\end{table}

\section{Results}{\label{results}}

\subsection{Spontaneous emergence of collective aster rotation}
Aster motility is a result of symmetry breaking in forces that arises from a combination of multiple forces: (i) kinesin-5 complexes that either zipper or separate asters based on the orientation of astral MTs, (ii) the bending forces from polymerizing MTs at the cell boundary, (iii) dynein pulling forces at the cell boundary (iv) stochasticity in astral MT lengths and (v) Brownian forces corresponding to the energy $k_BT$. However, none of these individually have any innate ability to drive directional motion due to the radial symmetry of asters and Brownian motion, as illustrated in Figure \ref{fig:asterschem}. We find the collective interactions result in three distinct form of aster patterns that depend on motors and MT dynamics: %of MTs with diffusible kinesin-5, boundary based pulling by cortical dyneins and noise originating from dynamic instability: 
(I) centering, (II) hexagonal lattice and (III) spontaneous rotation (Figure \ref{fig:minimalRotationSystem}). %The simulation images depict spatial patterns of asters at the end of 300 s, while their trajectories demonstrate motility. The trajectories are plotted over the last 30 s of the simulation for clarity.\\
Cortical dyneins exert an outward pulling %force on multiple confined asters, 
opposing the inward force generated by MT bending %absence of MT dynamics result in localize at steady state at the cell boundary, without any particular arrangement (Figure \ref{fig:minimalRotationSystem}a). The
and when combined with MT dynamics %on the other hand results in the 
result in the inward force dominating resulting in steady state {\it (I) centering} of asters (Figure \ref{fig:minimalRotationSystem}a). Even though MT lengths %$\langle L_{MT} \rangle$ of MTs 
are identical in dynamic and static cases, % as those of fixed-length MTs as determined by $f_c$, $f_r$, $v_g$ and $v_s$, the fluctuations in MT length 
the rare long filament bending against the cell membrane produces sufficient inward-forces to overcome the outward pulling due to dynein. % force of the cortical dyneins.\\
Kinesin-5 motors %simulated with asters without filament length dynamics 
results in a steady state {\it (II) hexagonal lattice} % with asters at their center. The structure 
resembling molecular crystal arrangements, as a result of kinesin-5 pushing forces on asters due to anti-parallel MTs (Figure \ref{fig:minimalRotationSystem}b). % generating pushing forces, but the structure is abolished if 
Dynamic instability abolishes these structures due to fluctuations in the overlap lengths.% The addition of `noise' through MT dynamic instability distorts this structure and results in aster separation without any particular order.\\
Cortical dyneins and diffusible kinesin-5 combined result in sliding motion along the circular cell boundary and coupling of asters respectively, which when combined with `noise' in form of MT dynamic instability break symmetry and result in steady-state {\it (III) spontaneous rotation}, analogous to `swarming' dynamics (Figure \ref{fig:minimalRotationSystem}c). % , driven by kinesin and dynein motors. %{\it Rotation:}   produces a well separated structure, comparable to the hexagonal packing that arises in presence of just kinesin-5 (Figure \ref{fig:minimalRotationSystem}c). The presence of cortical motors however produces short range back-and-forth 
Based on the role of dynein motors in force generation for the emergence of coherent aster rotational motion, we expected the motor density to be an important parameter and proceeded to test the systematic effect of motor density.

%----------------------------------------------------------------------------------------------------------------------------------
\begin{figure}[ht!]
	\begin{center}
		\includegraphics[width=0.7\textwidth]{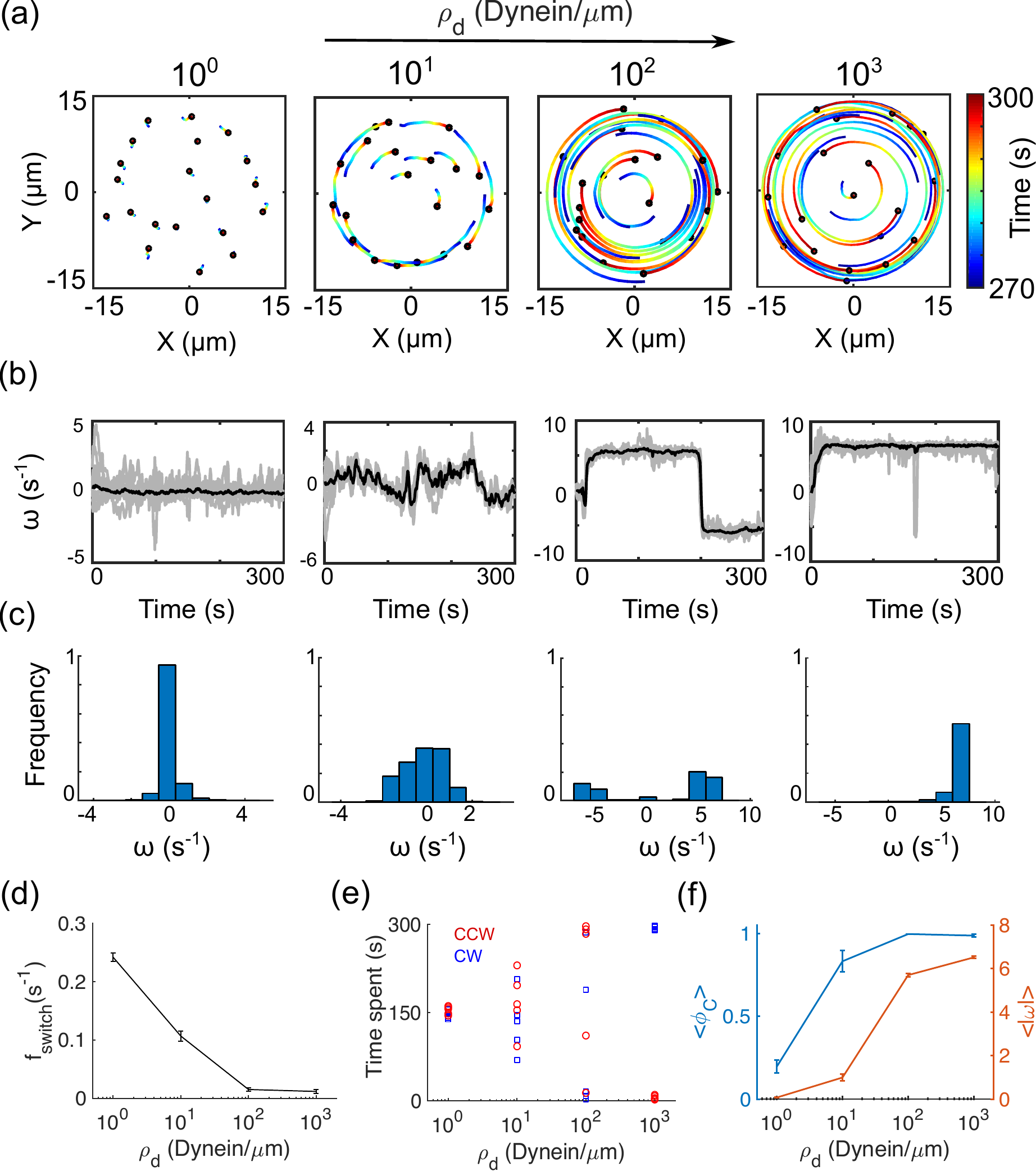}
		\caption{{\bf Onset of sustained aster rotation at threshold dynein density.} (a) Aster trajectories over the last 30 s are plotted for varying densities of cortical dynein ($\rho_{d}$= 10$^{0}$ to 10$^{3}$ motors/$\mu$m). Colorbar: time. (b) The instantaneous angular velocity ($\omega$) for each aster (gray) and the mean angular velocity (black) in time is plotted for increasing $\rho_{d}$. (c) Represents the frequency distribution of ($\omega$) for the corresponding simulations. (d) The switching frequency  and (e) total time spent in either clockwise or counter-clockwise direction is plotted for increasing densities of dynein over the entire trajectory. (f) The mean coherence in rotation ($\phi_{C}$) and the mean angular velocity ($\langle | \omega | \rangle$) over the last 30 s is plotted as a function of increasing dynein density. Cell radius: 15 $\mu$m, $N_{a}$=20, surface density of kinesin is ($\rho_{k}$= 10$^{10}$ motors/$\mu$m$^{2}$) and diffusion coefficient ($d_{d}$) of dyneins is 10 $\mu$m$^{2}$/s, J = -0.3 $\mu$m/s. N$_{runs} = 5$. Error bars: standard error.}
		\label{fig:dyndens}
	\end{center}
\end{figure}

%----------------------------------------------------------------------------------------------------------------------------------

\subsection{Dynein density dependent onset of rotation patterns}
The localization of cortical dyneins on a circular boundary that walk on MTs is expected to result in a `gliding' transport of filaments, with a strong rotational component. However, due to the radial geometry of asters, MTs from the same aster have been shown to encounter antangoistic forces resulting in a tug-of-war \cite{athale2014}. In our model, stochasticity in opposing forces resulting from MT length fluctuations produce an element of randomness that is expected to transiently resolve the tug of war. %MT transport and cytoplasmic kinesin-5 form the minimal system for onset of collective sliding and rotation movements. We systematically 
We find increasing the density of dynein ($\rho_{d}$) results in increasingly sustained rotation (Figure \ref{fig:dyndens}(a)), confirming the central role for dynein force generation when combined with stochastic MTs and kinesin-5 motors. The time traces of instantaneous angular velocity ($\omega$) exhibit are smaller in magnitude and fluctuate more in presence of few motors, while increasing numbers of dyneins result in higher amplitudes and fewer variations in angular velocity (Figure \ref{fig:dyndens}(b)). %The traces of individual asters undergo short and rapid fluctuations in time at low dynein density. With increasing densities the fluctuations decrease while the magnitude of angular velocity increases. 
%The instantaneous velocity, $\omega$  traces from individual asters exhibit fluctuations in time that decrease with increasing $\rho_{d}$ along with an 
Increasing synchronization between the individual traces is observed for increased motor densities. The pooled frequency distribution of  $\omega$ at low dynein density is distributed sharply around zero, increases in spread for $\rho_d$ of 10 motors/$\mu$m,  is bi-modal for 100 motors/$\mu$m and is biased to one side at a high density of $10^3$ motors/$\mu$m, indicative of persistent motion (Figure \ref{fig:dyndens}(c)). The switching frequency, $f_{switch}$, is calculated as the total number of switch events per unit time and quantifies the persistence in the direction. Consistent with the trend in $\omega$ we find increasing $\rho_d$ results in decreased $f_{switch}$ (Figure \ref{fig:dyndens}(d)). Rotating asters are not chiral since the time spent by individual asters moving in clockwise (CW) and counter clockwise (CCW) orientations is equal between multiple iterations. Individual simulations result in a spontaneous choice of an orientation and the collective movement is entrained, with no particular preference for CW or CCW motion (Figure \ref{fig:dyndens}(e)). However, due to the shorter sliding events at low densities, the proportion of time spent in either state is comparable which begins to diverge for increasing dynein densities.\\

%----------------------------------------------------------------------------------------------------------------------------------

\begin{figure}[ht!]
	\begin{center}
		\includegraphics[width=0.7\textwidth]{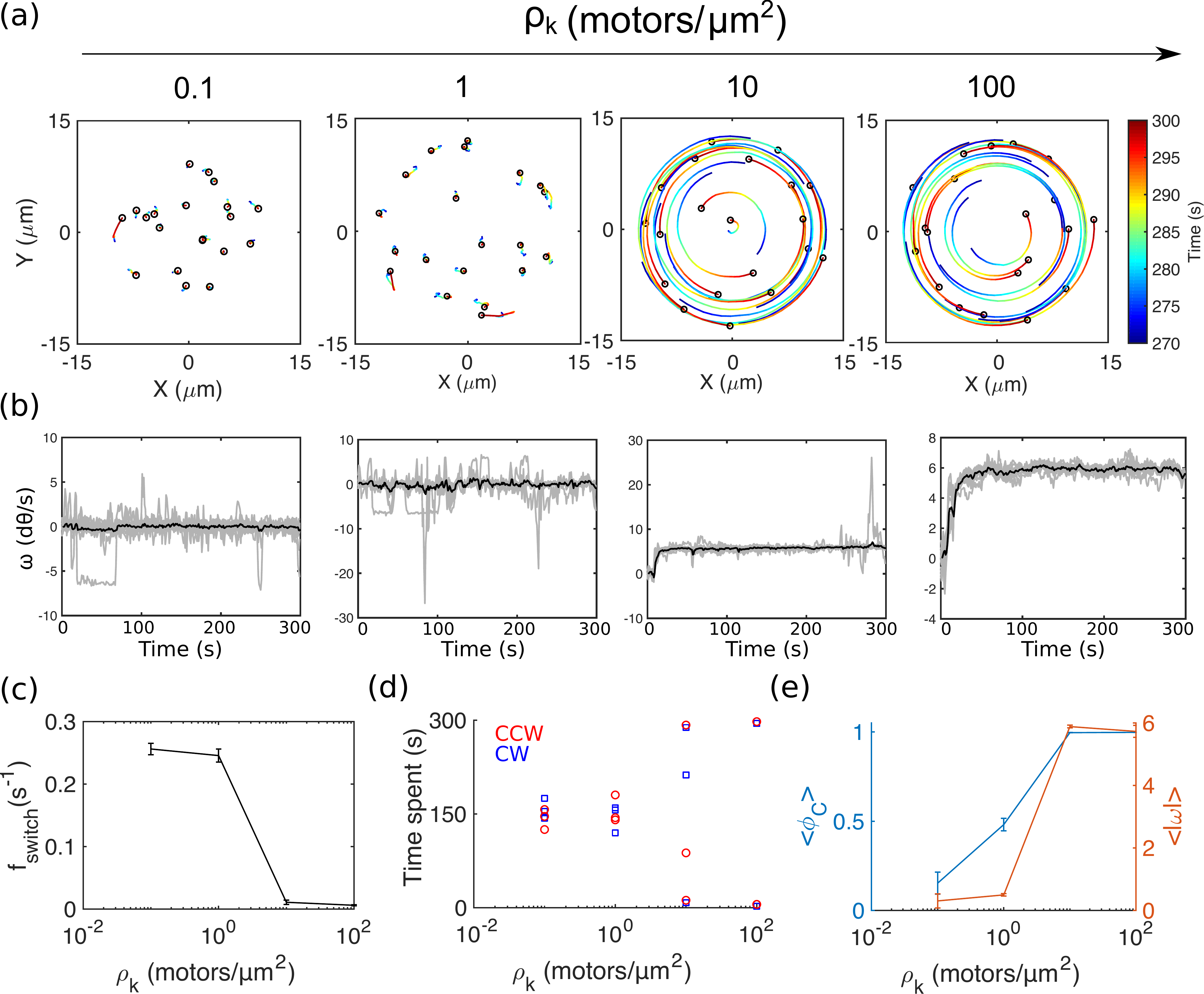}
		\caption{{\bf Strength of inter-aster interactions dictate onset of sustained collective rotation.}(a) 
			Representative XY trajectories of asters for varying densities of kinesin-5 motors in the cytoplasm is plotted over the last 30 s. Color encodes for the trajectory time. (b) The instantaneous angular velocity, $\omega$ is plotted in time. Gray curves represent individual aster track while the ensemble mean is represented in black. 
			(c) The switching frequency over the entire trajectory is plotted for varying kinesin-5 densities on (X-axis, log scale). (d) The total time spent in either clockwise (blue) or counter-clockwise (red) direction is plotted for increasing kinesin-5 density over the entire trajectory. (e) The measure of coherence in rotation, $\phi_{c}$ averaged over the last 30 s is plotted over the varying values of kinesin-5 density ( X axis- on the log scale) on the left y-axis and the ensemble mean over the last 30 s for the angular velocity is plotted on the right y-axis. $\rho_{d}= 10^2$ motors/$\mu$m, $\rho_{a}$ = 0.03 asters/$\mu$m$^{2}$, N$_{runs}=$3. Error bar indicates SEM.}
		\label{fig:ScanKin5den}
	\end{center}
\end{figure}
%----------------------------------------------------------------------------------------------------------------------------------

In order to quantify the emergent collective order in a system of rotating asters, we measure a rotational order parameter $\phi_{R}$ comparable to previous reports for cells \cite{lober2015} and active particles \cite{alaimo2016}. $\phi_{R}(t)$ is the mean rotational order parameter at time $t$ over $N$ asters expressed as:
\begin{equation}
	\phi_{R}(t)  =  \frac{1}{N} \cdot \sum\limits_{i=1}^N \hat{e}_{{\theta}_{i}}(t) \cdot \hat{v_{i}}(t) 
	\label{eq:R}
\end{equation}
Here, $\hat{v}$ is the unit velocity vector and $\hat{e}_{{\theta}_{i}}$ is the unit angular direction vector of the aster. The dot product of these two terms is averaged over the total number of asters (particles) $N$ where $i$ is the index of each aster (particle). $\phi_{R}$ can take values ranging between -1 for counter clock wise motion (CCW) and +1 for clock-wise motion, while zero indicates the absence of rotation. Since there is no preference observed for CW or CCW motion, the modulus of the rotational order averaged over time and between multiple asters measures the coherence of motion ($\phi_{C}$) as described by the expression:
\begin{equation}{\label{eq:CohROP}}
	\phi_{C} = \sum\limits_{t=t_{s}}^T  \langle| \phi_{R}(t) | \rangle 
	\label{eq:ROP}
\end{equation}
The time average is taken at steady state between $t_{s}$ the time for onset of rotation (typically 270 s) and  total time $T$ (300 s). % since sustained rotations which approximate steady state are seen in the time between 270 and 300 s. 
%The time average of $\phi_{R}(t)$  
We observe the measure of rotational coherence increases with $\rho_{d}$ and saturates at high dynein densities, which is mirrored by the mean angular velocity over the last 30 s transitioning from low to high value at a $\rho_{d}$ of $\sim$ 100 motors/$\mu$m (Figure \ref{fig:dyndens}(f)). Taken together, it suggests that a minimal number of dyneins (10 dyneins/$\mu$m) is sufficient for the onset of rotation while a sustained, coherent and persistent rotation emerge only at a higher dynein density (100 dyneins/$\mu$m). While cortical dynein is expected to play a role in the onset of rotation, kinesin-5 motors are thought to entrain the collective transport. In order to examine the importance of kinesin in the ordered rotational motion of asters, we test the effect of their density.

%----------------------------------------------------------------------------------------------------------------
% Effect of MT flux on coherence in rotation
\begin{figure}[ht!]
	\begin{center}
		\includegraphics[width=0.65\textwidth]{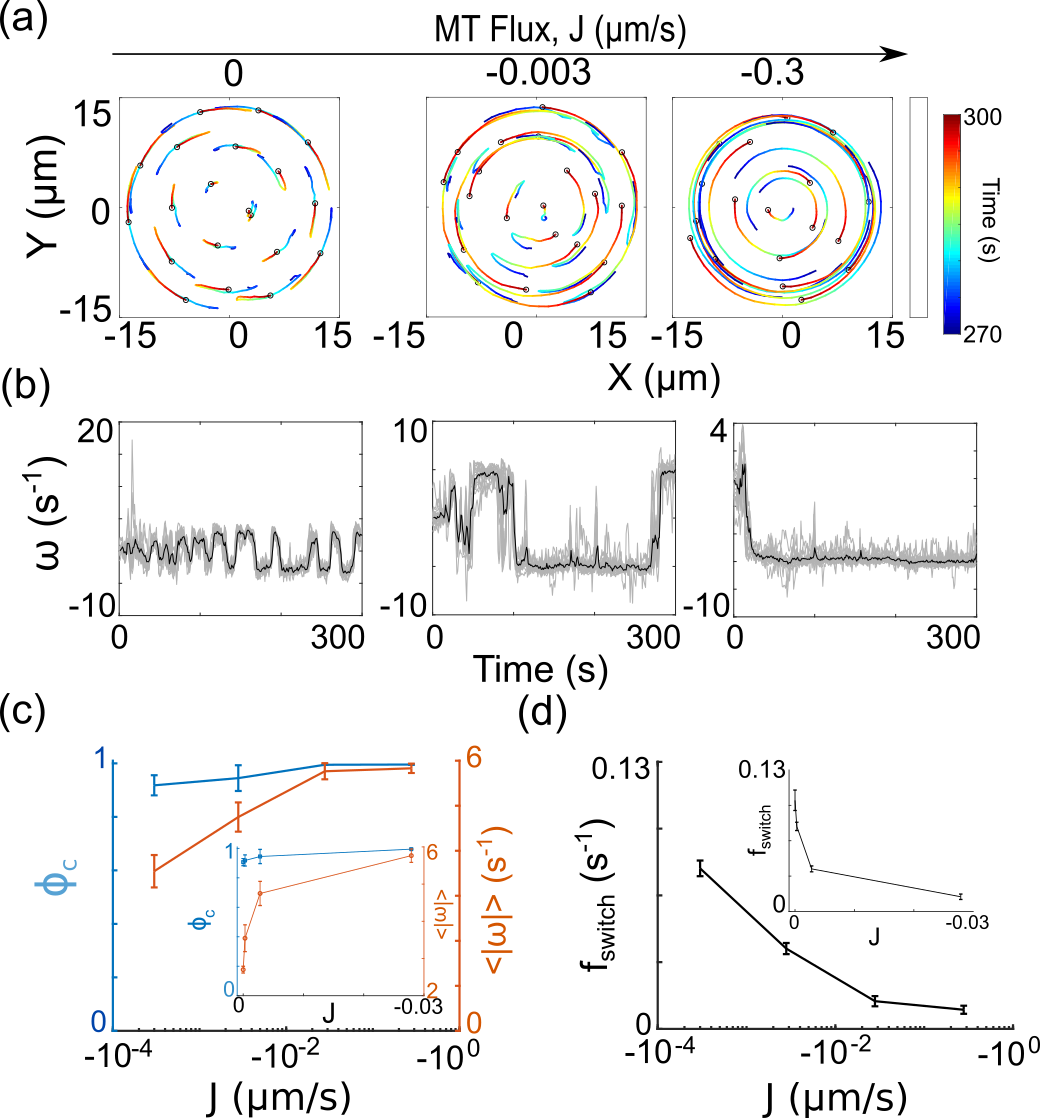}
		\caption{{\bf Effect of MT length fluctuations on persistence of rotation.} (a) The aster trajectories over the last 30 s is plotted for increasing values of MT flux (noise). Color encodes for the trajectory time. (b) The instantaneous angular velocity, $\omega$ is plotted in time. Gray curves represent individual aster track while the ensemble mean is represented in black. (c) The measure of coherence in rotation, $\phi_{c}$ averaged over the last 30 s is plotted over the varying flux values (log scale) on the left y-axis and the ensemble mean over the last 30 s for the angular velocity is plotted on the right y-axis. {\it Inset:} Linear plot. (d) The switching frequency over the entire trajectory averaged for n=3 runs ($\pm$s.d.) is plotted against the MT flux rate J (x-axis, log scale). The inset represents the same data with a linear x-axis. J was varied by modifying the frequencies of catastrophe ($f_{c}$) and rescue ($f_{r}$). Here, $\rho_{d} =10^2$ motors/$\mu$m, $\rho_{k}$ = 10 motors/$\mu$m$^{2}$, N$_{a}$ =20. }
		\label{fig:Scanmtflux}
	\end{center}
\end{figure}
%----------------------------------------------------------------------------------------------------------------

\subsection{Kinesin-5 numbers dictate strength of inter-aster interactions }

% I need to provide motivation and background for these
Kinesin-5 motors are local coupling factors between asters that produce a segregating force generated in the absence of dynein and MT dynamic instability (Figure \ref{fig:minimalRotationSystem}). Thus their role in entraining collective rotational motility is not necessarily intuitive, other than as a mechanism of exerting local coupling forces. Locally, kinesin-5 complexes bound to a pair of MTs from neighboring asters can generate pushing or pulling forces between the asters based on the orientation of the astral MTs to which they are bound- anti-parallel MTs result in asters being pushed apart or segregated, while parallel MTs result in asters being pulled together or `zippered' (Figure \ref{fig:asterschem}(d)). %Based on the motor polarity and MT orientation it is expected that an. A pair of cytoplasmic aster in a multi-aster scenario experiences multiple push-pull forces at any instant due to the many such interacting pairs of asters. 
%Thus these multiple combinatorial interactions result in an effective noise in the system and the strength of the noise can be varied by modulating the kinesin-5 numbers. 
Interestingly a low density of kinesin-5 fails to produce any sign of rotational motion, and only above a threshold motor density does collective rotation emerge (Figure \ref{fig:ScanKin5den}(a)). This is also evidenced by the low mean angular velocity $\sim 0$ below the threshold density (Figure \ref{fig:ScanKin5den}(b)) and the frequent switching due to only local fluctuations (Figure \ref{fig:ScanKin5den}(c)). Only once the threshold density has been crossed to we observe a choice of either CW or CCW rotation (Figure \ref{fig:ScanKin5den}(d)) with a sudden jump in the mean angular velocity and increase and saturation of the measure of rotational coherence (Figure \ref{fig:ScanKin5den}(e)). % at and beyond the density of 10 motors/$\mu$m$^{2}$. 
%However, at this threshold density the diffusion coefficient of kinesin-5 did not effect the collective rotation patterns. This suggests kinesin-5 density determines the strength of aster-aster interactions. 
This vital role of coupling kinesin motors, is comparable to the local coupling introduced in SPP models. Here, astral MT overlaps are maintained only when sufficient numbers of kinesin-5 motors continue to remain bound at steady state. This appears to be essential for the onset of coherent multi-aster rotation. 

The maintenance of local coupling through kinesin depends heavily on a minimal overlap of filaments between pairs of asters. While aster density can ensure the proximity of MTs, the filaments will need to be stable for sufficient time for kinesin-5 motors to bind and walk. We therefore proceeded to test the sensitivity of the model to the dynamicity of MT lengths.

%----------------------------------------------------------------------------------------------------------------   
\begin{figure}[ht!]
	\begin{center}
		\includegraphics[width=0.6\textwidth]{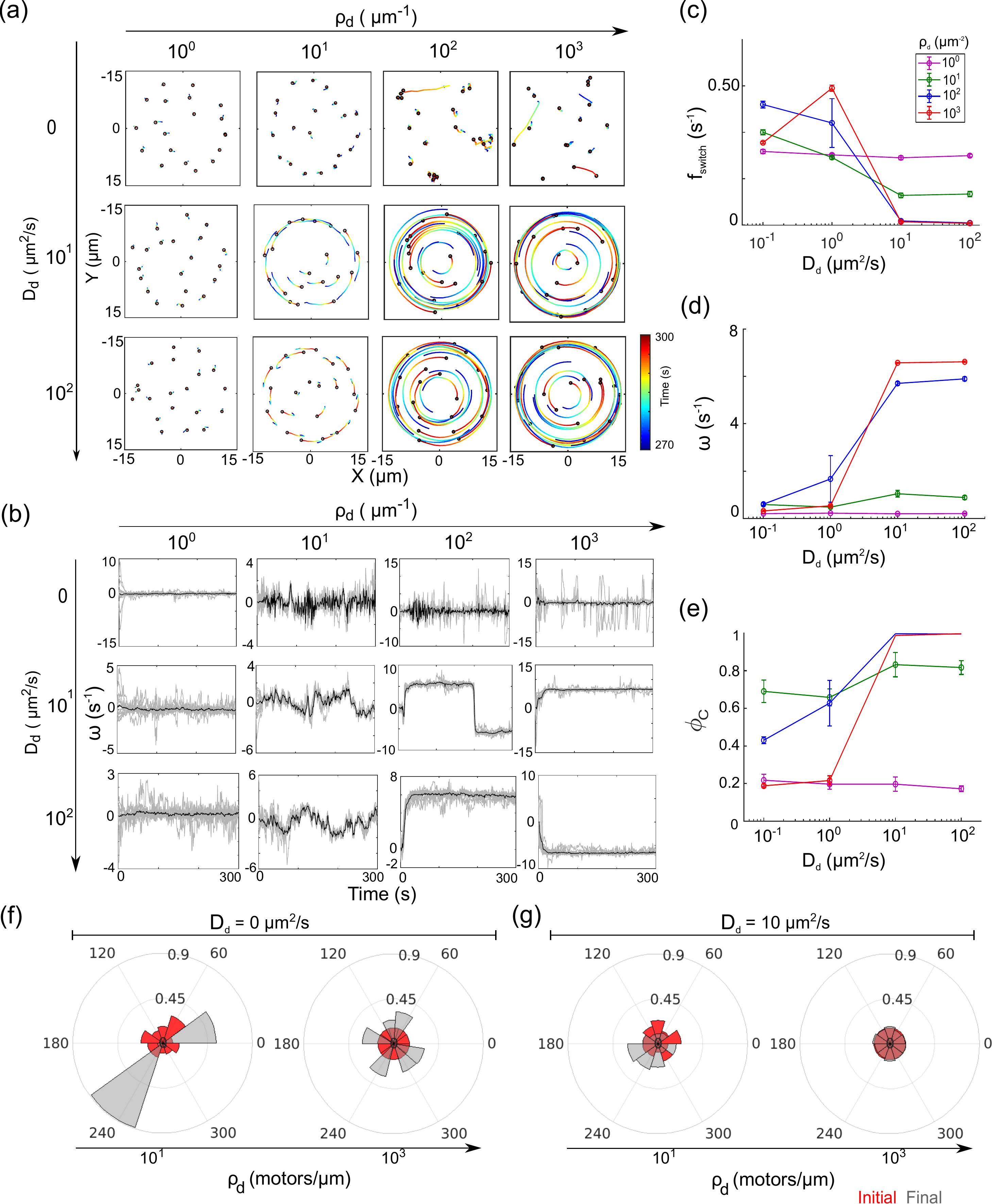}
		\caption{{\bf Diffusion of dynein restores the uniform redistribution at the cortex essential for the sustained rotation.} (a) The aster trajectories over the last 30 s and (b) the instantaneous angular velocity ($\omega$) in time is plotted for varying dynein diffusivity, $D_{d}$ {\it (column)} and dynein density, $\rho_{d}$ {\it (row)}. (c) The switch frequency over the entire trajectory is plotted as a function of dynein diffusion coefficient. Colors indicate varying dynein densities. The ensemble average over last 30 s for (d) mean angular velocity ($\langle | \omega | \rangle$) and (e) mean coherence measure ($\phi_{C}$) is plotted for increasing values of dynein diffusion coefficient,($D_{d}$). Distribution of dyneins at the cortex from a single run at the start (red) and end (gray)of the simulation for low and high dynein density is represented for (f) low and (g) high dynein diffusivity. N$_{a}$ = 20, J = -0.3 $\mu$m/s, $\rho_k =10$ motors/$\mu$m$^2$. N$_{runs}$ = 5. Error bars indicate SEM.}
		\label{fig:RotationDiffcoeff}
	\end{center}
\end{figure}
%----------------------------------------------------------------------------------------------------------------

\subsection{Coherence of rotation and the effect of MT length fluctuations}
MT dynamics {\it in vivo} differ between interphase when the flux $J>0$ corresponding to state of unbounded growth and slow dynamics, and mitosis when MTs are in a bounded state with $J<0$ and fluctuations in MT lengths \cite{verde1992}. The transition between these states has been attributed to the presence of microtubule associated proteins (MAPs) that bind to and modulate the dynamics of MTs, primarily by regulating the transition rates between growing and shrinking stages. Flux also varies across organisms, which could arise from intrinsic differences in tubulin polymerization or evolutionary differences in MAPs. 

We capture this difference in our model by modulating the flux rates and examine their effect on the patterns. By varying only the frequencies of catastrophe $f_c$ and rescue $f_r$ while keeping mean length $<L>$ and velocity of growth $v_{g}$ and shrinkage $v_{s}$ constant, we solved Equations \ref{eq:mtdynlen} and \ref{eq:mtdynflux} to obtain a range of values for $J$. MT dynamics is expected to influence the stability of overlaps of pairs of MTs, the instantaneous number of motors that can bind to MTs and fluctuations in the bending energy. To our surprise, {\it increasing} the magnitude of flux from $0$ to % $-3\times 10^{-3}$ and 
$-3\times 10^{-1}$ resulted in more persistent rotation of asters (Figure \ref{fig:Scanmtflux}(a)). This is confirmed by the {\it reduced fluctuation} in the angular velocity at steady state (Figure \ref{fig:Scanmtflux}(b)) and rapid increase and saturation of the mean rotational coherence $\phi_c$ and angular velocity $<|\omega|>$ variables for increasing flux (Figure \ref{fig:Scanmtflux}(c)). The decrease of switching frequency ($f_{switch}$) with increasing magnitude of $J$ further confirms the role that MT dynamics appears to play in reinforcing coherent motion (Figure \ref{fig:Scanmtflux}(d)). This feature of our model of increase in rotational order as a function of increasing noise diverges from the Viscek-type where increasing `noise' decreases the order of collective motion \cite{vicsek1995}. The difference could relate to the nature of `noise', since in our model it only indirectly affects mobility, while in SPP models `noise' increases the randomness of particle motion.

A higher degree of MT length flux likely results in higher turnover of binding events, and a more equal distribution of cortical force generators, dyneins. This is the likely cause for the increased coherence in the system. Another means by which cortical forces could be redistributed is the mobility of dynein. As a result we proceed to test whether the diffusive mobility of cortical dyneins plays a role in the collective transport of asters. %Increase in flux, increases the encounters at the cell cortex and with the kinesin-5 in the cytoplasm. It is expected that 
%The expected decrease in the duration of MT-motor interactions with increased flux, with a converse increase in frequency of interactions,  increases thereby resulting in an increased collective rotational order. While, the order in collective motion of active particles increase with particle density in the Viscek like models consistent also observed here for increasing dynein and kinesin-5 densities. In the Viscek models, the movement of active particles is determined based on the velocity and direction of the neighbouring particles. Here, interactions of an aster with its neighbor is governed by the activity of the kinesin-5 motors and pulling by the cortical dyneins that are present uniformly in the space. %This could be explained by the fact that rotation onset is stochastic, and increased flux improves the changes of rotation, while MT-motor interactions are optimal. 

%----------------------------------------------------------------------------------------------------------------

\begin{figure}[ht!]
	\begin{center}
		\includegraphics[width=0.7\textwidth]{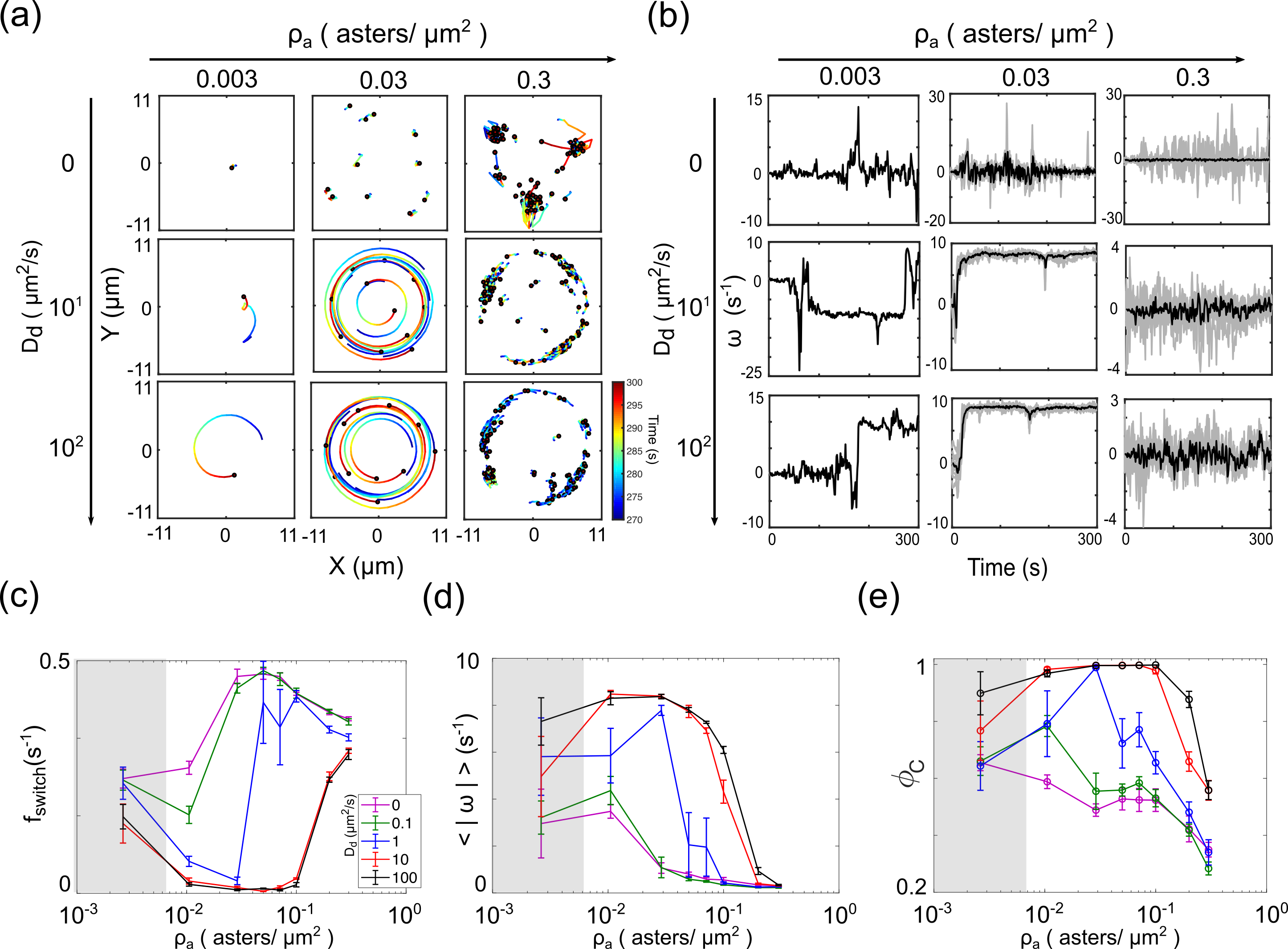}
		\caption{{\bf Aster density dependent multi-aster motility patterns.} (a) The aster trajectories over the last 30 s and (b) instantaneous angular velocity ($\omega$) in time is plotted for varying aster density, $\rho_{a}$ {\it (column)} and dynein diffusion coefficient, $D_{d}$ {\it (row)}. (c) The switch frequency over the entire trajectory is plotted as a function of aster density. Colors indicate dynein diffusion coefficient. The ensemble average over last 30 s for (d) angular velocity  and (e) measure of coherence in rotation is plotted for increasing values of aster density (X-axis, log scale). The gray shaded area in (c-e) corresponds to data with 1 aster in the cell. $\rho_{d} = 10^2$ motors/$\mu$m, $\rho_{k}$ = 10 motors/$\mu$m$^{2}$, J = -0.3 $\mu$m/s. N$_{runs}$ = 5. }
		% (f,g,h) The motility patterns across aster densities and dynein diffusivity is broadly categorized into 3 distinct types as depicted in contour plots for the data in (c,d,e) respectively.
		\label{fig:asterden}
	\end{center}
\end{figure} 
%----------------------------------------------------------------------------------------------------------------

\subsection{Dynein diffusion limits symmetry breaking}
%{Symmetry breaking is diffusion limited} 
% An alternate story would be to have just the walking motor in the previous section. Here based on the schematic as the motivation for having moving dyneins - introdue that and then discuss the results here.
%We next asked whether an inhomogeneity in the distribution of cortical dyneins could effect the symmetry breaking and onset of collective rotation of the asters. 
Inside cells, a fraction of dynein is bound to the cell cortex in dynamic clusters with a rapid turnover due to unbinding and diffusion \cite{dehmelt2014}. % is observed to detach from the cortex, diffuse in the cytoplasm and attach elsewhere, undergo MT mediated displacement and drag along the cortex, experience drag due to the actin flows and undergo movements along the cortex due to binding on the lipid membrane. Thus 
We model this mobility of dyneins through diffusion of the motors along the cortical region determined by an effective diffusion coefficient of dynein ($D_{d}$). In order to test whether this diffusive mobility has any effect on the aster-mobility, we tested the effect of varying $D_{d}$ in a simulated cell with optimal kinesin-5 density and cell size. In continuation with the idea that a minimal density of dynein is required for the onset of rotations, we also varied $\rho_d$, the density of dynein. %Since fluctuations drive the symmetry breaking we attempted to implement the dynein movement along the cortex by varying $D_{d}$. 
We find, decreasing dynein diffusivity from 10 to 0 $\mu$m$^{2}$/s resulted in complete abolition of rotational motion even when $\rho_{d}$  was greater than the threshold density required for rotation (Figure \ref{fig:RotationDiffcoeff}(a)). In other words, the presence of dynein alone is not sufficient to drive rotational motion, but also requires diffusive redistribution. Increasing $D_{d}$ above the threshold resulted in a saturation of coordinated motility as quantified by angular velocity (Figure \ref{fig:RotationDiffcoeff}(b)), with the obvious absence of dynein abrogating aster motility altogether. We observe a diffusion  ($D_{d}$) dependent phase-transition like behaviour in the switching frequency (Figure \ref{fig:RotationDiffcoeff}(c)), the steady state angular velocity (Figure \ref{fig:RotationDiffcoeff}(d)) and the coherence in rotational order parameter $\phi_C$ (Figure \ref{fig:RotationDiffcoeff}(e)). %The variability in $\phi_C$ between runs is the greatest at an intermediate value of 1 $\mu$m$^2$/s, as the system transitions from static asters to sustained rotation. The similar diffusion dependent trend is observed at higher densities of dynein. 
%At low dynein densities $D_{d}$ has no effect, consistent with our previous results demonstrating the essential nature of dyneins in the aster rotation. %and asters continue to be static suggesting both density %motility ($D_{d}$ = 10 $\mu$m$^{2}$/s) and mobility of dyneins is essential to sustain coherent aster rotations that emerge at and beyond the optimal dynein density. 

This diffusion-limited behaviour of asters arises from the uniformity of free dyneins. % restored above a threshold value of $D_d$, lost due to the effective drag experienced by the motors. 
When $D_{d}$ is below the threshold, dyneins do not adequately redistribute when unbound from MTs, resulting in formation of clusters irrespective of dynein density (Figure \ref{fig:RotationDiffcoeff}(f)) while diffusive dynein results in a steady state unform distribution (Figure \ref{fig:RotationDiffcoeff}(g)). %At and above optimal dynein density $D_{d}$, uniform presence of dyneins and rotations is thus sustained. 
Taken together, it suggests that homogeneity in dynein distribution at the cortex is essential for uniform pulling that can sustain sliding and drive collective motion. Since both kinesin and dynein act to transport the MT asters, we proceeded to ask whether aster density is likely to play a major role in collective motility, based on the predicted role for particle density in SPP models . %As a result we proceed to test whether aster density per cell plays a role in the patterns. %It is expected for fewer asters in the system, reduced number of dyneins with low mobility rates should be able to drive collective rotation and for higher number of asters both the dynein numbers and the mobility should be higher in order to generate uniform pulling forces.

%----------------------------------------------------------------------------------------------------------------
\begin{figure}[ht!]
\begin{center}
	\includegraphics[width=0.5\textwidth]{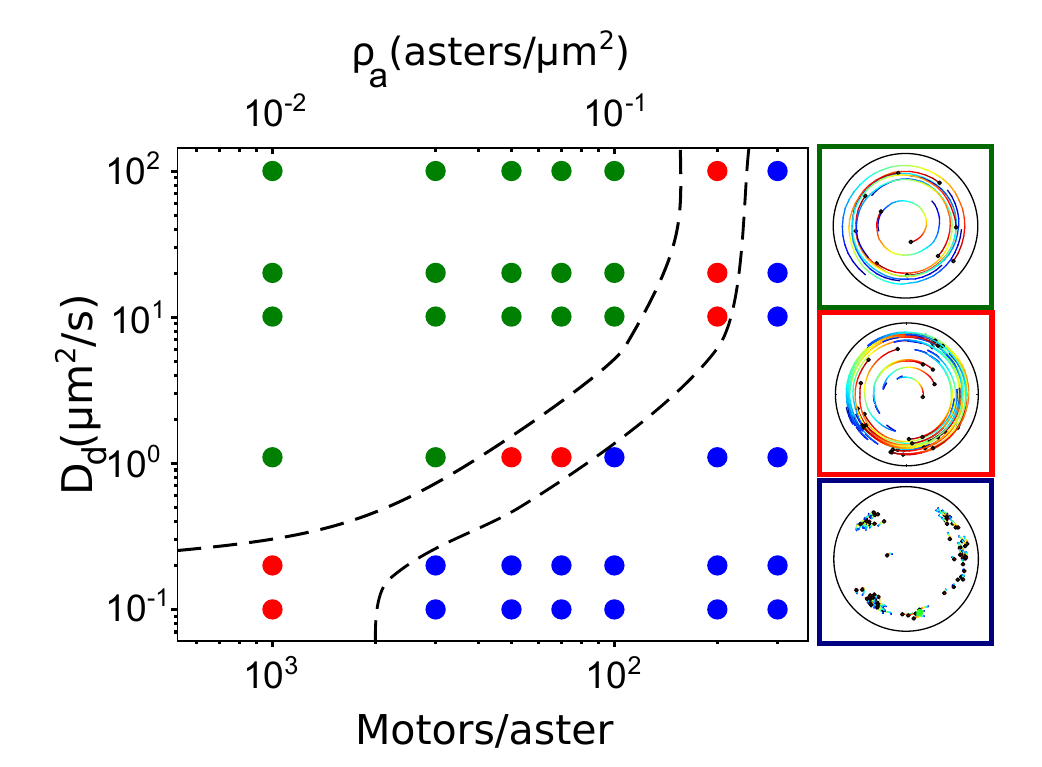}
	\caption{{\bf Motility patterns dependent on motor numbers and dynein distribution.} The effect of aster density $\rho_a$ and dynein diffusion coefficient $D_d$ was based on the calculated average angular velocity ($\omega$), coherence in rotation ($\phi_c$) and frequency of switching ($f_s$) into three clusters (Figure \ref{fig:clustersRaw}). The three clusters correspond to qualitatively distinct forms of mobility: sustained, coherent and persistent rotation {\it(green)}, coherent rotation with variable persistence {\it(red)} and a lack of rotation {\it(blue)}, as seen in the representative XY trajectories. The dashed-line is a guide to the eye separating the clusters. Cell size R = 11 $\mu$m, cortical dynein density is 100 motors/$\mu$m and cytoplasmic kinesin-5 density is 10 motors/$\mu$m$^{2}$.}
\label{fig:summaryphase}
\end{center}
\end{figure}
%----------------------------------------------------------------------------------------------------------------

\subsection{Critical density of asters required for rotation}
%{An optimal aster density required for onset of rotational}
In SPP models, collective ordered transport of particles emerges when the strength of local coupling and particle density are both optimal. %The strength of aster-aster interactions in addition to the activity of kinesin-5 and MT dynamics can be modulated by varying the aster density ($\rho_{a}$). 
%In Viscek models, the movement of the active particle is determined by the particles within it's interaction zone and the boundary conditions. 
In our multi-aster systems, while asters mechanically interact with the boundary, for coupling we require kinesin-5 activity while dynein acts at the boundary. %The movement of the asters is determined by the interactions with these motors and the MT dynamics which effect their encounters. Thus at constant conditions of motors, aster numbers were varied to see the combined effects of motors and MT dynamics on the interaction and transport of asters thereby the effect on collective rotation. 
%In a cell of 11 $\mu$m radius with optimal values of MT dynamics, dynein density ($\rho_{d}$ of 10$^{2}$ motors/$\mu$m$^{2}$) and effective diffusivity ($D_{d}$ of 10 $\mu$m$^{2}$/s) as in 
Rotation onset was observed only at an optimal value of aster density when $N_A$ were varied from 1 to 114, keeping all other conditions optimal for coherent aster rotations (Figure \ref{fig:asterden} (a,b), middle row). Both high and low aster asters $\rho_a$ failed to produce rotation. On the other hand, due to the diffusion limitation of dynein when $D_{d}$ was varied we found the patterns did not change above a threshold value of dynein diffusion. % While, on decreasing the $D_{d}$ rotation is observed at low aster densities diminishing to static asters in the absence of $D_{d}$. The range increases with increasing
Indeed for $D_{d}$ $>$ 1 $\mu$m$^{2}$/s, the system undergoes collective rotation for an optimal range of aster density, attaining minimal switching transitions (Figure \ref{fig:asterden}(c)) and maximal velocity (Figure \ref{fig:asterden}(d)) and rotational coherence (Figure \ref{fig:asterden}(e)). %Additionally, the variability between runs reduces at the optimal density of around $10^{-1}$ aster per $\mu$m$^2$. 
The observed density dependence deviates from the  kinetic phase transitions in the SPP models, which we understand as resulting from % The drop beyond the optimal density in the system could be simply due to the non-linear increase in forces with aster numbers which is 
insufficient number of force generators required to drive large-scale rotation at high densities, and too few particle interactions at low densities. This is consistent with restoration of rotation when the dynein density is increased by ten fold (data not shown).  Thus, we observe that collective rotation is limited by aster density, where the low density effects arise from lack of coupling, while at high densities the relative number of motors %. The symmetry is broken 
per asters play a role. %and sustained motility only emerges when the forces exerted by the motors are (a) optimal and symmetric in space and (b) MT-motor interactions are frequent.

Our model therefore predicts a complex set of components and behaviour that are predicted to result in coherent, collective rotational transport of asters involving density of motors and asters, the stochasticity of MTs and is diffusion-limited in terms of cortical dynein.

%Thus our model predicts the combination of a uniform outward force from cortical motors, separating activity of kinesin and noise due to dynamic instability can produce spontaneous rotations in multiple MT asters in a confined system. The multi-asters exhibit phase transition like trend for increasing dynein density and MT dynamics that saturates at optimal range. While, dependence on aster density is different, optimal only for a narrow range limited by the number of motors per astral MTs in the cell which effects the pulling and {\textcolor{blue}{re-distribution of dyneins at the cortex.}}

%----------------------------------------------------------------------------------------------------------------
\begin{figure}[ht!]
	\begin{center}
		\includegraphics[width=0.7\textwidth]{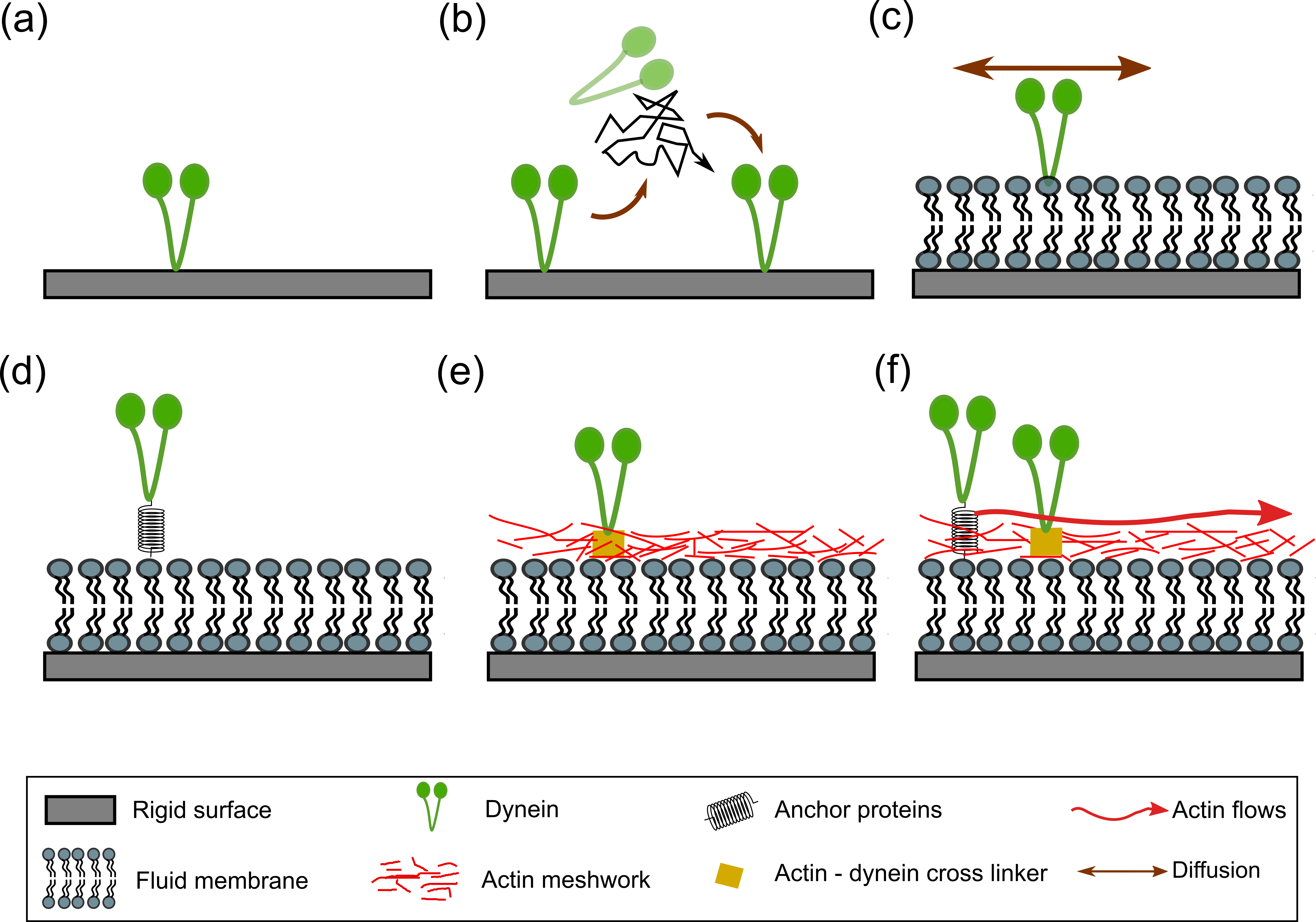}
		\caption{{\bf Determinants of cortical dynein mobility.} Dynein (green) mobility at the cortex can be the result of multiple forms of anchorage: (a) immobilized on a rigid cortex (grey), (b) unbind (arrow), diffuse in the cytoplasm and bind at another location (arrow), (c) diffuse in a lipid membrane (bidirectional arrow), (d) coupled to a stretchable linker (spring) that is embedded in the membrane, (e) attached via an adapter protein (box) that cross-link with the actin cortex or (f) linker or actin cross-linker are actively transported by actin flows (red arrow).}
		\label{fig:MotorCortex_Motivation_V2}
	\end{center}
\end{figure}
%----------------------------------------------------------------------------------------------------------------

\section{Discussion and Conclusions}

The emergence of self-organized multi-aster swarming in confinement is comparable to a wide range of biological systems that span several scales- from molecules, through bacterial populations to large animal swarms. The specific properties of molecular motors and their critical role in cell physiology, growth and division makes their study particularly important %{\textcolor{blue}{NK: how about pertinent instead of important?}}. 
The role of radial MT arrays seen here in the spontaneous emergence of patterns depends critically on four factors: (a) rigid boundary resulting in bending of MTs, (b) diffusively redistributed cortical dynein forces that generate a tug-of-war of MTs which when resolved produces circumferential movement, (c) forces of local coalescence and separation of asters driven by kinesin-5 like motors producing coupling and (d) stochasticity due to filament polymerization kinetics that break the symmetry of the system. %Depending upon the parameter space, different motility patterns emerge that can be broadly classified as coherent, persistent and transient sliding events that do not undergo collective rotation. 
Dynein at the cortex is critical for sustained rotation with kinesin-5 resulting in local coupling that enhances coherence of rotation. Counter-intuitively increased MT lengths increases rotational persistence, in effect due to increased `noise'. A critical density of asters is required for the persistence of steady-state rotation, with both low- and high-density limits resulting in a loss of rotation. %Collectively, the average number of motors interacting per aster determines the effect of length fluctuations on multi aster movement.

% Self organized model independent of biochemistry 
In order to summarize the range of aster density and diffusion limitation of cortical dynein in determining rotational onset, we cluster the angular velocity $\omega$, rotational coherence $\phi_c$ and frequency of switching $f_s$ into three clusters (Figure \ref{fig:clustersRaw}). We find this clustering produces three regions in the phase plane of dynein diffusion $D_d$ and aster density $\rho_a$ -- a region of sustained and persistent rotation, one completely lacking rotation and a transition zone between the two (Figure \ref{fig:summaryphase}). This picture of the onset of collective motion, dependent on density and noise resembles the SPP reported by Vicsek et al. \cite{vicsek2012}. However, we observe coherent rotation requires an {\it optimal} aster density, in contrast to the onset of ordered collective motility or swarming {\it above} a certain particle density. % {\textcolor{blue}{NK: If we comment on a collective density term i.e. number of motors interacting or available per aster then we can say that order increases with increasing density and thus I have replaced in most of the places. This then also falls in line with Viscek model. However the discrepncy is in our simulations with increase in noise - order increases in contrary to the Viscek type model. We could further extrapolate this in the following rational. In VM increase in noise would imply less time or degree to re-orient while that happens here for low MT dynamics. If we discuss noise and density - in these two aspects - we can map it to viscek and general flocking models. The analogy is not straightforward but we have three active particles here dynein, kinesin and MTs thus it would be best if we discuss and summarize in terms of collective terms which would then make sense too. Does this make sense?} 
This reversal in trends compared to SPP models could arise from the discrete nature of coupling in our model, inspired by the need for mechanistic detail compared to the simplifying implicit coupling in SPP models. In particular at high particle (aster) density, the kinesin-5 motors per aster become limiting, as depicted in Figure \ref{fig:summaryphase}. An additional difference to general SPP models is the diffusion-limitation of cortical dynein mobility, again arising from the discrete and physically realistic model of boundary forces. Thus our model we believe predicts that physically inspired details of collective transport by local coupling can also result in qualitatively different behaviour than that predicted by the general SPP models. The physical detail in our model has the advantage over a more abstract model of the potential to test our predictions in experiment.

%This can be understood to arise from the discrete nature of the force generators, in particular dynein. A similar effect is expectedly also observed when kinesin-5 diffusivity is minimized, due to clustering on MTs. % {\sout{Mixtures of microtubules (MT) and motor with an energy source, have been previously shown to show similar properties to such active systems and linear MT filaments were seen to self-organize to centered aster-like structures and transition to vortices depending on the system size and coupling motor densities}} 

%The theoretical basis of phase transitions in spatial pattern formation have been well described with detailed models \cite{Sankararaman:2004aa,surrey2001}, that help better understand experiments using mixtures of linear, polarized filaments and molecular motors as in the case of linear MTs mixed with motors \cite{nedelec1997,nedelec2001,Suzuki2017}. A density dependent phase transition and swirling of filaments has also been reported in actin-myosin systems which  \cite{schaller2010}, suggesting a relation to general SPP models. Our work demonstrates density-dependent rotational motion emerging from more complex radial MT arrays, that were expected to undergo tug of war and only local motion due to geometry .

The role of cortical dynein is critical in generating the rotational component of motion due to MT binding and dynein motility. Therefore, we believe it is important to consider the behavior of cortical dyneins in the context of reported {\it in-vivo} interactions in cells. For simplicity we consider the dynein to be bound to the cortical membrane throughout the simulation. Mobility of dynein is dependent on whether the motor is bound to MTs. When bound to MTs, the motors are dragged based on the motor stretch, independent of the motor stepping, while free motors undergo 1D diffusion in the cell boundary corresponding to the cortex. %The  effective diffusion coefficient and the . The cellular %{\sout{(i)}} 
The model of cortical dynein mobility is based on previous reports based on %{\textcolor{blue}{that suggest MT-dynein encounters at the cortex are short and frequent while dynein location is dynamic. In a study from}} 
dynamic microscopy of MT associated dynein speckles consisting of multiple molecules were found at the cell cortex with a high turnover over seconds time-scales  \cite{dehmelt2014}. In cells of the fission yeast {\it Schizosaccharomyces pombe}, cortical dyneins bound to MTs detach from the cortex and MTs, diffuse in cyotosol and eventually bind at an another cortical location \cite{vogel2009,vadynein2013}. Such mobility is distinct from {\it in vitro} gliding assays with motors immobilized on the glass cover-slip at anchored point (Figure \ref{fig:MotorCortex_Motivation_V2}(a)). Additionally, MTs have been seen to themselves mediate dynein localization and re-distribution at the cortex \cite{dehmelt2014,markus2011offloading,medema2014}, that has motivated models of redistribution of the motor at the cortex (Figure \ref{fig:MotorCortex_Motivation_V2}(b)). Additionally, {\it in vitro} MT-motor gliding assays with motors anchored in supported membrane bilayers suggest the collective transport properties change due to motor `slippage' and diffusion \cite{nelson2014myo, grover2016}. We the reported motor diffusion coefficients in lipids for our model of dynein diffusion (Figure \ref{fig:MotorCortex_Motivation_V2}(c)), based on reports for kinesins and myosins \cite{leduc2004,grover2016,nelson2014myo}. %However, lipid diffusion rates reported span over a wider range between 0.1 to 10 $\mu$m$^{2}$/s based on lipid composition and lipid model systems \cite{beckers2020} and also found to be different across the cell types \cite{greenberg1993}. 
Additionally, lipids mobility itself has been demonstrated to affect motor distribution as seen in case of kinesin motors that cluster along MTs by lateral diffusion of membrane lipids in a gliding assay, in the absence of motor activity \cite{lopes2019}. However, evidence from biochemical interactions and localization studies suggest dynein is bound to adaptor proteins which in turn are membrane bound such as Num1 and dynactin \cite{va2016}, mcp5 \cite{va2017mcp5} or ternary complexes such as NuMA/LGN/G$\alpha$i \cite{lee2012,kotak2019,ddncluster2018kiyomitsu}. Therefore a future improvement to models of cortical dynein mechanics would also need to take into account adaptor mechanics (Figure \ref{fig:MotorCortex_Motivation_V2}(d)). Actin networks that line the cell cortex, on the cytoplasmic side of the inner membrane of most animal cells, result in hindered diffusion of membrane proteins anchored in the plasma membrane, like for example GPCRs \cite{Kasai:2011aa,Kasai:2014aa}. The resulting `corralling' of receptors by an actin meshwork modifies the hopping probability resulting in local clustering and aggregation of receptors \cite{Deshpande:2017aa} and hindering the recruitment of dynein at the cortex \cite{huse2019}. Therefore the effect of spatial heterogeneity in dynein mobility could further improve our ability to predict localization patterns observed in specific cell types (Figure \ref{fig:MotorCortex_Motivation_V2}(e)). Actin-associated proteins %\cite{afadin2016}, huntingtin, MISP \cite{misp2013,misp2018} and myosin \cite{va2017mcp5} 
regulate the distribution of dynein anchors at cortex during spindle positioning in cells \cite{kotak2019, huse2019} and actin flows result in direct displacement of dyneins along the cortex in {\it C. elegans} embryos \cite{De-Simone:2018aa}, suggesting further details of cellular mechanics could allow for a direct comparison to {\it in vivo} experimental dynamics (Figure \ref{fig:MotorCortex_Motivation_V2}(f)). Integrating the insights obtained from {\it in vivo} single-molecule imaging, {\it in vitro} reconstitution of lipid-motor \cite{lopes2019} and lipid-actin \cite{invitroactinlipid2014} systems, will go a long way to improve the quantitative precision and qualitative match between the model predictions and experiments in order to provide insights into the role of cortical dynein MT aster positioning in cells.
%Taken together, the dynein localization and behaviour at the cortex is complex which we have modelled phenomenologically by allowing for the movement of dyneins. In future, aster reconstitution studies can be coupled to the lipid-motor (\cite{grover2016,lopes2019}) and lipid-actin (\cite{invitroactinlipid2014,petraactinlipid2013}) reconstitution studies which would 

% evolutionary implications
The absence of more widespread spontaneous rotation of asters in animal cells suggests that biological systems may have evolved mechanisms to suppress rotations. %Our results provide mechanistic insights on the observed interactions between astral MTs and cortical dyneins. %Dynein based cortical pulling is evolutionary conserved mechanism across cells during spindle positioning however, rotations are not often observed. 
In our model dynein must be diffusively redistributed at the cortex for sustained aster rotation to occur, in order to produce spatially homogeneous pulling forces throughout the cortex. Once motors are bound to MTs, they walk towards the minus end, resulting in a local clustering at multiple locations where MTs contact the cortex. Upon unbinding from the MT dynein diffusion results in dissipation of any clustering. On the other hand experimental studies from cells indicate cortical dyneins are localized multi-protein clusters at the cortex \cite{dehmelt2014,markus2011She,va2017mcp5,vadynein2013,ddncluster2018kiyomitsu}. Additional sources of spatial heterogeneity in dynein distribution could be the switching between states of activity by regulators as seen in budding yeast dynein activation and cortical localization by Num1 \cite{Sheeman:2003aa}. %(c) single motor properties such as binding-unbinding kinetics. In simulations, we observe loss of rotation in presence of clustered dyneins irrespective of the dynein density. 
 %In simulations, a uniform dynein density at the cortex results in onset of sliding events while a sustained collective aster rotation emerge at optimal densities. Secondly, 
Our model also requires lateral sliding of MTs along the cortex for coherent rotational motility to emerge. However, in experiments the nature of MT-motor interactions depend on cell shape, and may be end-on or lateral, in spherical or cylindrical cell geometries respectively % {\textcolor{blue}{correlated with cell-cycle stage - metaphase and anaphase}}
\cite{dumont2017}. In addition, dynein capture based shrinkage at the cortex is reported in yeast \cite{estrem2017} and T-cells \cite{jason2013} which may further act as brake to the sustained lateral interactions also discussed in \cite{evoGundersen2002}. % of  observed which seems to be dependent on the shape of the cell, stage of the cell-cycle and/or the accessory proteins that are engaged with MTs or dyneins. Typically, end-on interactions are observed during the centering processes while off-centering movements are driven by the lateral interactions. The geometry of the cell from round to a cylindrical shape tends to shift the MT-cortex interaction from end-on to the lateral  irrespective of the cytoplasmic state of the cell. While, stabilization of MTs at the onset of anaphase was reported to be essential for correct spindle positioning \cite{uhlmann2005}. MT stabilization implies slower MT dynamics. We observe transient back-and forth sliding events in the absence of MT dynamics and sustained persistent rotation with increase in MT dynamics in simulations. An increase in catastrophe rates is also observed for the MTs sliding along the cortex in budding yeasts at the anaphase \cite{carminati97,estrem2017}. Such behavior would result in length reduction of MTs and thereby suppress the lateral interactions. Additionally, the MT interaction at the cortex could be end on or lateral based on the anchor proteins to which it binds (\cite{ddncluster2018kiyomitsu}). 
Thus, multiple mechanisms responsible for precise positioning of spindles could have been an optimization to suppress the sustained lateral interaction of MTs at the cortex thereby preventing the rotations in the cell.

% experiments : known and future
Our study is distinct from the multiple reports of cytoskeletal filament vortices and sustained flows due to the MT geometry and multiplicity of asters. In contrast, cytoplasmic streaming in {\it Drosophila} oocytes driven by kinesin-dependent MT transport \cite{serbus2005} due to MTs sliding on other, cortically immobilized MTs \cite{lu2016}, requires linear MTs. More recently, a combination of experiment and theory have demonstrated large motor driven MT vortices in space on either 2D surfaces \cite{Sumino2012} and in 3D confined compartments with {\it Xenopus} extracts \cite{Suzuki2017}, both of which were performed using linear MTs. There is a clear lack of studies that examine collective properties of asters. {\it In vivo} MT asters during the first embryonic division of {\it Caenorhabditis elegans} are seen to oscillate due to pulling by cortical dyneins and a force asymmetry arising from MT bundling and motor density. While this has been studied in detail in experiment and simulation, the studies typically invoke only a pair of asters, coupled at the metaphase plate \cite{kozlowski2007}, but do not address the properties of multiple asters. In contrast, the centering ability of asters based on pushing at the cell membrane and pulling by cytosolic dynein motors seen in sea urchin embryos \cite{tanimoto2018}, suggests such cellular processes are robust to positioning errors arising from thermal noise. Since our predictions of spatial patterns are based on similar mechanistic components- filaments, diffusible and cortically anchored motors and a rigid boundary-arising from the presence of many asters and multiple, diffusible motors. The choice of parameters of motor and MT mechanics is based on experimentally measured values, with the aim of predicting the outcome of potential experiments. Indeed recent developments with linear MTs encapsulated in lipid droplets together with motors, demonstrate a transition from random to astral geometries dependent on collective mechanics \cite{bauman2014,juniper2018}. In future, the encapsulations of multiple asters could be used to test some of our model predictions.

% viscek system
The system described here is based on experimentally reported models and parameters of MTs, asters and the motor mechanics and kinetics of dynein and kinesin-5.  %While it is qualitatively comparable to the general model of density and noise-driven global ordered motion seen in locally coupled active particles \cite{vicsek1995}, the details of the system make it distinct. In particular, in Vicsek class of models collective ordered transport of the individual, self propelled objects can emerge when the strength of local coupling and `noise' or stochasticity are both optimal. The analogy of our detailed biological cytoskeleton-motor system can be formulated in terms of equilibrium model properties, where temperature can be compared to the random fluctuations in MT length and pressure compared to aster density. The optimal density effect of the system arises from boundary effects that dominate at high densities of asters. The breakdown in pattern is analogous to the density dependent swarming patterns of collectively migrating cells in experiment and theory \cite{szabo2006}. 
Our computational model demonstrates how local mechanical coupling combined with stochastic fluctuations and circular boundary effects can result in the emergence of coherent swarming motility of MT asters. The model demonstrates a diffusion limitation of motors, due to the discrete nature of the force generators. %In it's absence, both asters and dyneins aggregate locally around the cell boundary. 
This system transitions from `flocking' to `swirling' behaviour, depending on the mobility of the dyneins on the cell boundary. Thus we believe these results point to a novel biologically testable self-organized pattern forming system.

\section{Acknowledgements}
This work was supported by a grant from the Dept. of Biotechnology (DBT), Govt. of India BT/PR16591/BID/7/673/2016 to Chaitanya Athale. Fellowships from IISER Pune for integrated PhD, council for scientific research (CSIR) India 09/936(0128)/2015-EMR-1 and project assistantship from a DBT grant BT/PR16591/BID/7/673/2016 supported Neha Khetan. We acknowledge feedback about the biological motivation of the model from Thomas Lecuit.
%\end{acknowledgements}
%==========================

\section{Appendix}

%==========================
\section{Numerical simulations and data analysis}
Simulation code was based on C++ code of Cytosim, an OpenSource agent based simulation engine for cytoskeletal mechanics \cite{nedelec2007}. A typical simulation run with 20 asters and 7070 kinesin and 9500 dynein motors in a cell of radius 15 $\mu$m for 300 s required $\approx$ 200 minutes on a 12 core Intel machine (Xeon E5 2630) running Linux (Ubuntu 14.04) with 15.6 GB memory. All data analysis and plotting were performed in MATLAB R2017a (Mathworks Inc., USA). For clustering analysis, custom made Python scripts was written using scikit-learn package \cite{scikit-learn}. Heirarchial clustering (k-means) was performed using angular velocity, coherence order parameter and switching frequency as features. The silhouette score for 3 clusters of 0.758 was higher than for 2 or 4 clusters and used to determine the clusters in variable space (Figure \ref{fig:clustersRaw}). The labels of each cluster were then used to determine colour classes of aster motility (Figure \ref{fig:summaryphase}).

%% ========================= FIGURES

%% ==================================================================================
%\clearpage
%\newpage

%\appendix
%\section{}

\section{Supporting Information}
%{\label{sec:mtdynamics}}

\setcounter{table}{0}
\makeatletter 
\renewcommand{\thetable}{{\it SI} \@arabic\c@table} 

\begin{table}[ht!]
	\begin{center}
		\begin{tabular}{p{2cm} p{4.7cm} p{2.5cm} p{2.5cm}}
			\hline
			{\bf Parameter} & {\bf Description}       &   {\bf Value} & {\bf Reference}  \\    
			\hline 
			%		{\bf MT dynamic instability parameters:} & & & \\
			f$_{c}$         & Frequency of catastrophe  & 0.049 (s$^{-1}$), varied  & \cite{athale2008}, this study \\
			f$_{r}$         & Frequency of rescue       & 0.0048 (s$^{-1}$), varied & \cite{athale2008}, this study \\
			v$_{g}$         & Growth velocity           & 0.196 ($\mu$m/s)          & \cite{athale2008} \\
			v$_{s}$         & Shrinkage velocity        & 0.325 ($\mu$m/s)          & \cite{athale2008} \\
			$\kappa$        & MT bending modulus        & 20 (N/m$^{2}$)            & \cite{gittes1993} \\
			{\it Aster:}    &                           &         &              \\
			N$_{a}$         &  Number of asters         &     40, varied  &  This study      \\
			N$_{MT}$        &  Number of MTs/aster      &     20  &  This study      \\
			\\     	
		\end{tabular}
	\end{center}
	\caption{{\bf MT parameters.} The microtubule and aster parameters used in simulations were taken from experimental measurements reported in literature, and where missing estimated.}
	\label{tab:mtsimparams}
\end{table}

\begin{table}[ht!]
	\caption{{\bf Transition frequencies and MT flux rates.} The parameters of MT dynamic instability: frequency of catastrophe (f$_{c}$, s$^{-1}$) and frequency of rescue (f$_{r}$, s$^{-1}$) was varied to obtain different MT flux rates (J) at constant value of growth velocity (v$_{g}$ = 0.196 $\mu$m/s), shrinkage velocity (v$_{s}$ = 0.325 $\mu$m/s) and mean MT length (L$_{MT}$ = 4.25 $\mu$m). The values were reported previously in \cite{verde1992,athale2008}.}
	\label{tab:mtDynparams}
	\begin{center}
		\begin{tabular}{p{2.5cm} p{2.5cm} p{2.5cm}}
			\hline     
			{\bf f$_{c}$ (s$^{-1}$)} & {\bf f$_{r}$ (s$^{-1}$)} &  {\bf J ($\mu$m/s)}   \\   
			\hline
			{\it Stabilized:} & & \\
			None        &			   None             &           0						 \\
			{\it Dynamic:} & & \\
			20.2683        &      33.5317           &        -0.0003         \\
			2.0527         &       3.3273           &        -0.0028         \\
			0.2312         &       0.3068           &        -0.0279         \\
			0.049          &       0.0048	          &        -0.2785         \\
			\hline
		\end{tabular}
	\end{center}
\end{table}

\clearpage
\newpage

\section{Supplemental Figures}
%%% Supporting figures
\setcounter{figure}{0}
\makeatletter 
\renewcommand{\figurename}{Figure}
\renewcommand{\thefigure}{S\@arabic\c@figure} 

\begin{figure}[ht!]
	\begin{center}
		\includegraphics[width=0.7\textwidth]{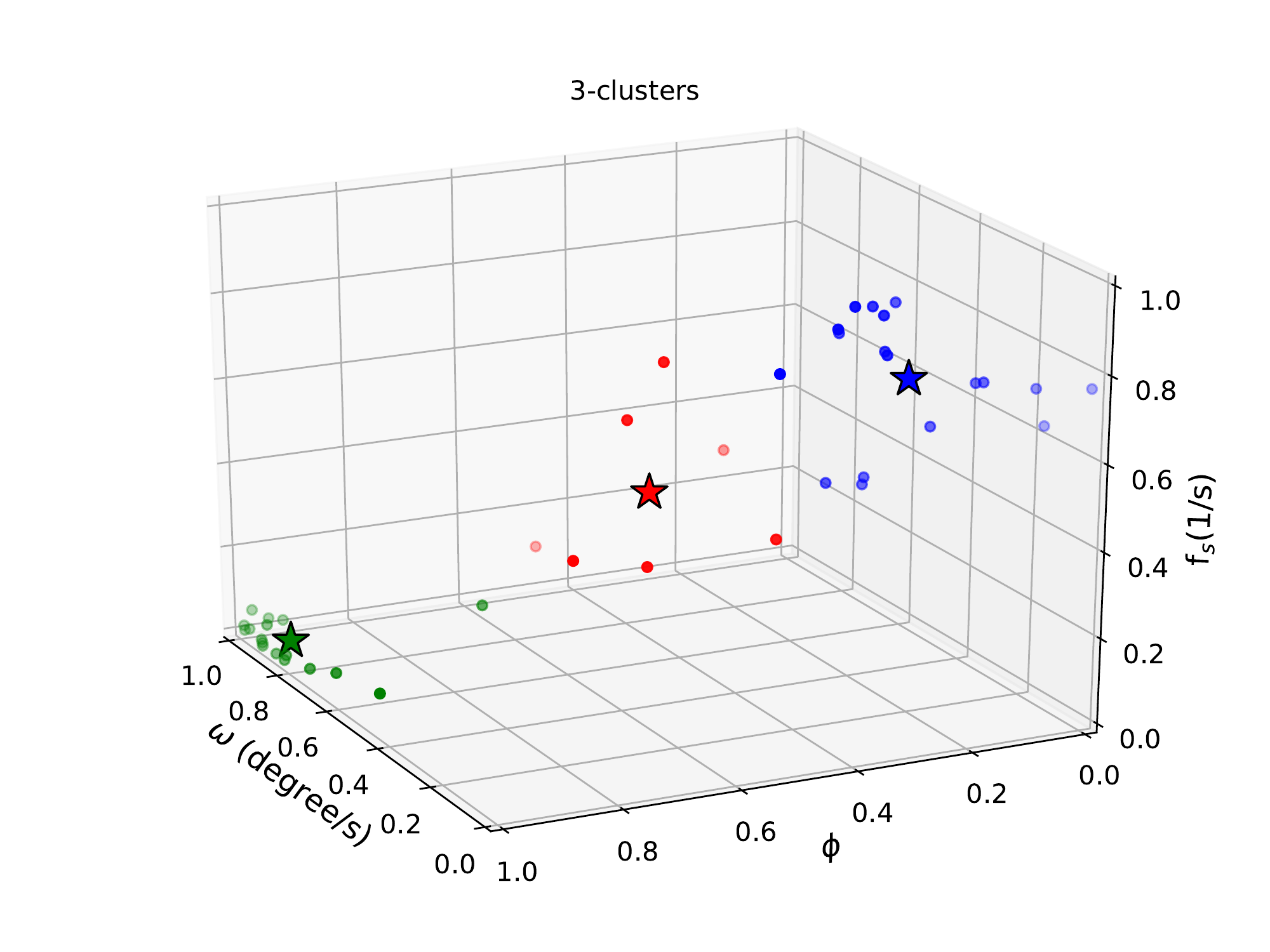}	
		\caption{{\bf Clusters of motility patterns for varying number of motors per aster and degree of uniformity at the cortex.} The parameter space for varying aster density and dynein diffusion coefficient was segregated into clusters using k-means clustering based on the measures of average angular velocity ($\omega$), coherence in rotation ($\phi_c$) and frequency of switching ($f_s$). The circles correspond to points in the parameter space and color represents the clusters: {\it(green)}
		 sustained, {\it(red)} coherent and persistent rotation and {\it(blue)} for coherent rotation with variable persistence and a lack of sustained rotation . The 'stars' indicate the centroid of the clusters. Cell size R = 11 $\mu$m, cortical dynein density = 100 motors/$\mu$ and cytoplasmic kinesin-5 density = 10 motors/$\mu$m$^{2}$.}
		\label{fig:clustersRaw}
	\end{center}
\end{figure}

\section{Supplemental Videos}
%%% VIDEOS
\setcounter{figure}{0}
\makeatletter 
\renewcommand{\figurename}{Video}
\renewcommand{\thefigure}{SV\@arabic\c@figure}

\begin{figure}[ht!]
	\begin{center}
		\includegraphics[width=0.4\textwidth]{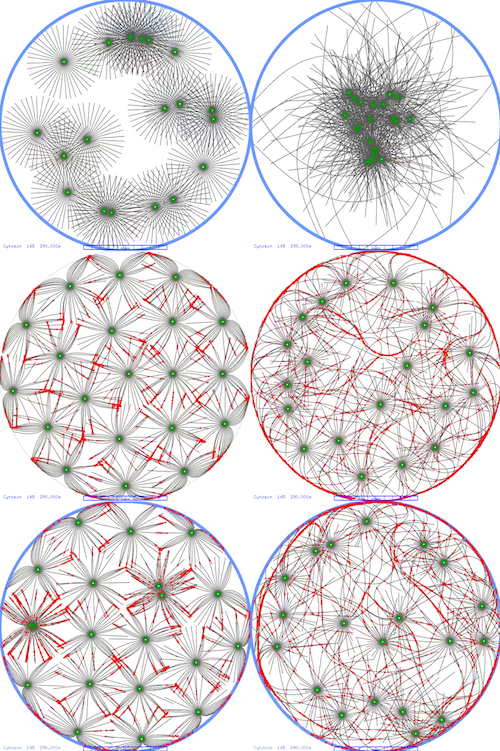}	
		\caption{The effect of {\it (top row)} cortical force generators, {\it (mid-row)} kinesin complexes and {\it (bottom row)} both kinds of motors on MT asters with the two columns simulating MTs lengths as {\it (left)} fixed {\it (right)} dynamically unstable. The video corresponds to results depicted in Figure 1. Time is 300 s, $N_a$ = 20, $\rho_d$ = 100 motors/$\mu$m, $\rho_k$ = 10 motors/$\mu$m$^2$, J = 0 and - 0.3 $\mu$m/s for stabilized and dynamic MTs.}
		\label{vid:patterns}
	\end{center}
\end{figure}
%Figure \ref{fig:minimalRotationSystem}

%==== REFERENCES
%\section{References}
\bibliography{multiasterpatterns_v4}
\bibliographystyle{plain}%apsrev4-2}

\end{document}